\documentclass[twocolumn]{aastex7}
\usepackage{appendix}
\usepackage{graphicx}
\usepackage{CJKutf8}
\usepackage{longtable} 
\usepackage{booktabs} 
\usepackage{chngcntr}
\counterwithin{figure}{section}
\usepackage[utf8]{inputenc}
\usepackage{comment}
\graphicspath{{./}{figures/}}

\newcommand{\Msun}{${\rm M}_{\odot}$}
\newcommand{\kms}{km\,s$^{-1}$}

\newcommand{\SII}{S~{\sc ii}}
\newcommand{\NII}{N~{\sc ii}}

\newcommand{\NaI}{Na~{\sc i}~D}

\newcommand{\CrII}{Cr~{\sc ii}}

\newcommand{\OII}{O~{\sc ii}}
\newcommand{\OIII}{O~{\sc iii}}

\newcommand{\FeIII}{Fe~{\sc iii}}
\newcommand{\HeI}{He~{\sc i}}
\newcommand{\NiII}{Ni~{\sc ii}}
\newcommand{\FeII}{Fe~{\sc ii}}
\newcommand{\CaII}{Ca~{\sc ii}}
\newcommand{\MgI}{Mg~{\sc i}}

\shorttitle{SN 2022erq}
\shortauthors{Zhai et al.}

\begin{document}

\title{SN 2022erq: A Superluminous Thermonuclear Supernova with Escalating Preexplosion Mass Loss}

\author[0009-0002-3956-6143]{Qian Zhai}
\affiliation{International Centre of Supernovae (ICESUN), Yunnan Key Laboratory of Supernova Research, Yunnan Observatories, Chinese Academy of Sciences (CAS), Kunming, 650216, China}
\email{zhaiqian@ynao.ac.cn}  

\author[0000-0002-8296-2590]{Jujia Zhang*}
\affiliation{International Centre of Supernovae (ICESUN), Yunnan Key Laboratory of Supernova Research, Yunnan Observatories, Chinese Academy of Sciences (CAS), Kunming, 650216, China}
\email[show]{* jujia@ynao.ac.cn}  

\author[0000-0002-7334-2357]{Xiaofeng Wang} 
\affiliation{Physics Department, Tsinghua University, Beijing 100084, China}
\email{wang\_xf@mail.tsinghua.edu.cn}

\author[0000-0002-7259-4624]{A. Pastorello}
\affiliation{INAF - Osservatorio Astronomico di Padova, Vicolo dell'Osservatorio 5, 35122 Padova, Italy}
\email{andrea.pastorello@inaf.it}

\author[0000-0003-3460-0103]{Alexei V. Filippenko}
\affiliation{Department of Astronomy, University of California, Berkeley, CA 94720-3411, USA}
\email{}

\author[0000-0001-8764-7832]{J\'ozsef Vink\'o}
\affiliation{HUN-REN CSFK Konkoly Observatory, MTA Centre of Excellence,
Konkoly Thege M. \'ut 15-17, Budapest, 1121, Hungary}
\affiliation{Department of Experimental Physics, Institute of Physics, University of Szeged, D\'om t\'er 9, Szeged, 6720 Hungary}
\affiliation{Department of Astronomy, University of Texas at Austin,
2515 Speedway, Stop C1400, Austin, TX, 78712-1205, USA}
\email{vinko.jozsef@csfk.hun-ren.hu}

\author[0000-0001-5955-2502]{Thomas G. Brink}
\affiliation{Department of Astronomy, University of California, Berkeley, CA 94720-3411, USA}
\email{}

\author[0000-0002-2547-0434]{Yunkun Han}
\affiliation{International Centre of Supernovae (ICESUN), Yunnan Key Laboratory of Supernova Research, Yunnan Observatories, Chinese Academy of Sciences (CAS), Kunming, 650216, China}
\email{}

\author[0000-0002-3334-4585]{G. Valerin}
\affiliation{INAF - Osservatorio Astronomico di Padova, Vicolo dell'Osservatorio 5, 35122 Padova, Italy}
\email{giorgio.valerin@inaf.it}

\author[0000-0002-1381-9125]{N. Elias Rosa}
\affiliation{INAF -- Osservatorio Astronomico di Padova, Vicolo dell'Osservatorio 5, 35122 Padova, Italy}
\affiliation{Institute of Space Sciences (ICE, CSIC), Campus UAB, Carrer de Can Magrans, s/n, E-08193 Barcelona, Spain}
\email{nancy.elias@inaf.it}

\author[0000-0002-7714-493X]{Yongzhi Cai} \affiliation{International Centre of Supernovae (ICESUN), Yunnan Key Laboratory of Supernova Research, Yunnan Observatories, Chinese Academy of Sciences (CAS), Kunming, 650216, China}
\affiliation{INAF -- Osservatorio Astronomico di Padova, Vicolo dell'Osservatorio 5, 35122 Padova, Italy}
\email{caiyongzhi789@gmail.com}

\author{Weili Lin}
\affiliation{Department of Astronomy, Xiamen University, Xiamen, Fujian 361005, China}
\email{}

\author[0000-0002-6535-8500]{Yi Yang}
\affiliation{Physics Department, Tsinghua University, Beijing 100084, China}
\email{}

\author[0000-0002-2636-6508]{WeiKang Zheng}
\affiliation{Department of Astronomy, University of California, Berkeley, CA 94720-3411, USA}
\email{}

\author[0000-0001-5316-2298]{Xiangcun Meng} 
\affiliation{International Centre of Supernovae (ICESUN), Yunnan Key Laboratory of Supernova Research, Yunnan Observatories, Chinese Academy of Sciences (CAS), Kunming, 650216, China}
\email{}

\author[0000-0002-2452-551X]{Chengyuan Wu}
\affiliation{International Centre of Supernovae (ICESUN), Yunnan Key Laboratory of Supernova Research, Yunnan Observatories, Chinese Academy of Sciences (CAS), Kunming, 650216, China}
\email{}

\author[0009-0003-3758-0598]{Liping Li} \affiliation{International Centre of Supernovae (ICESUN), Yunnan Key Laboratory of Supernova Research, Yunnan Observatories, Chinese Academy of Sciences (CAS), Kunming, 650216, China}
\email{}

\author[0009-0005-2963-7245]{Zeyi Zhao}
\affiliation{International Centre of Supernovae (ICESUN), Yunnan Key Laboratory of Supernova Research, Yunnan Observatories, Chinese Academy of Sciences (CAS), Kunming, 650216, China}
\email{zhaozeyi@ynao.ac.cn}

\author[0000-0003-1349-6538]{J. Craig Wheeler}
\affiliation{Department of Astronomy, University of Texas at Austin,
2515 Speedway, Stop C1400, Austin, TX, 78712-1205, USA}
\email{wheel@astro.as.utexas.edu}

\author[0000-0003-1072-2712]{Jose L. Prieto}
\affiliation{Instituto de Estudios Astrof\'isicos, Facultad de Ingenier\'ia y Ciencias, Universidad Diego Portales, Avenida Ej\'ercito Libertador 441, Santiago, Chile}
\email{jose.prietok@mail.udp.cl}

\author[0009-0000-0314-6273]{Jialian Liu}
\affiliation{Physics Department, Tsinghua University, Beijing 100084, China}
\email{}

\author{Gaici Li} \affiliation{Physics Department, Tsinghua University, Beijing 100084, China}
\email{}
\author[0009-0004-4256-1209]{Shengyu Yan} \affiliation{Physics Department, Tsinghua University, Beijing 100084, China}
\email{}
\author[0009-0001-0660-1902]{Fangzhou Guo}
\affiliation{Physics Department, Tsinghua University, Beijing 100084, China}
\email{}

\author[0000-0003-2375-2064]{C. P. Guti\'errez}
\affiliation{Institut d'Estudis Espacials de Catalunya (IEEC), 08860 Castelldefels (Barcelona), Spain}
\affiliation{Institute of Space Sciences (ICE, CSIC), Campus UAB, Carrer de Can Magrans, s/n, E-08193 Barcelona, Spain}
\email{cgutierrez@ice.csic.es}

\author[0000-0001-8257-3512]{E. Kankare}
\affiliation{Tuorla Observatory, Department of Physics and Astronomy, University of Turku, FI-20014 Turku, Finland}
\email{erkki.kankare@utu.fi}

\author[0000-0002-3664-8082]{Peter Lundqvist}
\affiliation{The Oskar Klein Centre, Department of Astronomy, Stockholm University, AlbaNova, SE-10691, Stockholm, Sweden}
\email{peter@astro.su.se}

\author[0000-0001-5221-0243]{S. Moran}
\affiliation{School of Physics and Astronomy, University of Leicester, University Road, Leicester LE1 7RH, UK}
\email{shane.moran@leicester.ac.uk}

\author[0000-0003-4254-2724]{A. Reguitti}
\affiliation{INAF - Osservatorio Astronomico di Padova, Vicolo dell'Osservatorio 5, 35122 Padova, Italy}
\affiliation{INAF - Osservatorio Astronomico di Brera, Via E. Bianchi 46, 23807 Merate (LC), Italy}
\email{andrea.reguitti@inaf.it}

\author[0000-0003-1450-0869]{I. Salmaso}
\affiliation{INAF - Osservatorio Astronomico di Capodimonte, Salita Moiariello 16, 80131 Napoli, Italy}
\affiliation{INAF - Osservatorio Astronomico di Padova, Vicolo dell'Osservatorio 5, 35122 Padova, Italy}
\email{irene.salmaso@inaf.it}

\author[0000-0002-5571-1833]{M. D. Stritzinger}
\affiliation{Department of Physics and Astronomy, Aarhus University, Ny Munkegade 120, DK-8000 Aarhus C, Denmark}
\email{max@phys.au.dk}

\author[0000-0001-8178-0202]{S. Williams}
\affiliation{Tuorla Observatory, Department of Physics and Astronomy, University of Turku, FI-20014 Turku, Finland}
\affiliation{Finnish Centre for Astronomy with ESO (FINCA), Quantum, Vesilinnantie 5, University of Turku, FI-20014 Turku, Finland}
\email{scwilliams1987@gmail.com}  

\author[0009-0009-8633-8582]{J.-W. Zhao}
\affiliation{South-Western Institute for Astronomy Research, Yunnan Key Laboratory of Survey Science, Yunnan University, Kunming, Yunnan 650500, China}
\email{zhaojiewei@stu.ynu.edu.cn}

\author[0000-0002-4314-5686]{D.-D Shi}
\affiliation{Center for Fundamental Physics, School of Mechanics and opticelectrical Physics, Anhui University of Science and Technology, Huainan, Anhui 232001, China}
\email{ddshi@aust.edu.cn}

\author[0000-0002-0349-7839]{Jianrong Shi}
\affiliation{National Astronomical Observatories, Chinese Academy of Sciences, Beijing 100101, China}
\email{sjr@bao.ac.cn}

\author[0009-0000-7773-553X]{Z.-H. Peng}
\affiliation{School of Electronic Science and Engineering, Chongqing University of Posts and Telecommunications, Chongqing 400065, China}
\email{s230601017@stu.cqupt.edu.cn}

\author[0000-0002-3876-6330]{Tengfei Song}
\affiliation{International Centre of Supernovae (ICESUN), Yunnan Key Laboratory of Supernova Research, Yunnan Observatories, Chinese Academy of Sciences (CAS), Kunming, 650216, China}
\email{stf@ynao.ac.cn}  

\author[]{Yongyuan Xiang}
\affiliation{International Centre of Supernovae (ICESUN), Yunnan Key Laboratory of Supernova Research, Yunnan Observatories, Chinese Academy of Sciences (CAS), Kunming, 650216, China}
\email{xiangyy@ynao.ac.cn}  

\author[0000-0002-8585-4544]{Attila B\'odi}   \affiliation{HUN-REN CSFK Konkoly Observatory, MTA Centre of Excellence,
Konkoly Thege M. \'ut 15-17, Budapest, 1121, Hungary}
\affiliation{Department of Astrophysical Sciences, Princeton University,
Peyton Hall, 4 Ivy Lane, Princeton University, Princeton, NJ 08544}
\email{}

\author[0000-0002-6497-8863]{Borb\'ala Cseh}   \affiliation{HUN-REN CSFK Konkoly Observatory, MTA Centre of Excellence,
Konkoly Thege M. \'ut 15-17, Budapest, 1121, Hungary}
\affiliation{MTA-ELTE Lend{\"u}let "Momentum" Milky Way Research Group, Hungary}
\email{}

\author[0009-0008-2052-8474]{\'Agoston Horti-D\'avid}     \affiliation{ELTE E\"otv\"os Lor\'and University, Institute of Physics and Astronomy, 
P\'azm\'any P\'eter s\'et\'any 1A, Budapest 1117, Hungary}
\affiliation{HUN-REN CSFK Konkoly Observatory, MTA Centre of Excellence,
Konkoly Thege M. \'ut 15-17, Budapest, 1121, Hungary}
\email{}

\author[0000-0002-1663-0707]{Csilla Kalup}   \affiliation{ELTE E\"otv\"os Lor\'and University, Institute of Physics and Astronomy, 
P\'azm\'any P\'eter s\'et\'any 1A, Budapest 1117, Hungary}
\affiliation{HUN-REN CSFK Konkoly Observatory, MTA Centre of Excellence,
Konkoly Thege M. \'ut 15-17, Budapest, 1121, Hungary}
\email{}

\author[0000-0002-1792-546X]{Levente Kriskovics}    \affiliation{HUN-REN CSFK Konkoly Observatory, MTA Centre of Excellence,
Konkoly Thege M. \'ut 15-17, Budapest, 1121, Hungary}
\email{}

\author[0000-0002-8770-6764]{R\'eka K\"onyves-T\'oth}   \affiliation{HUN-REN CSFK Konkoly Observatory, MTA Centre of Excellence,
Konkoly Thege M. \'ut 15-17, Budapest, 1121, Hungary}
\affiliation{Department of Experimental Physics, Institute of Physics, 
University of Szeged, D\'om t\'er 9, Szeged, 6720 Hungary}
\email{}

\author[0000-0001-5449-2467]{Andr\'as P\'al}    \affiliation{HUN-REN CSFK Konkoly Observatory, MTA Centre of Excellence,
Konkoly Thege M. \'ut 15-17, Budapest, 1121, Hungary}
\email{}

\author[0000-0002-3658-2175]{B\'alint Seli}   \affiliation{ELTE E\"otv\"os Lor\'and University, Institute of Physics and Astronomy, 
P\'azm\'any P\'eter s\'et\'any 1A, Budapest 1117, Hungary}
\affiliation{HUN-REN CSFK Konkoly Observatory, MTA Centre of Excellence,
Konkoly Thege M. \'ut 15-17, Budapest, 1121, Hungary}
\email{}

\author[0000-0001-7806-2883]{\'Ad\'am S\'odor}   \affiliation{HUN-REN CSFK Konkoly Observatory, MTA Centre of Excellence,
Konkoly Thege M. \'ut 15-17, Budapest, 1121, Hungary}
\email{}

\author[0000-0002-1698-605X]{R\'obert Szak\'ats}   \affiliation{HUN-REN CSFK Konkoly Observatory, MTA Centre of Excellence,
Konkoly Thege M. \'ut 15-17, Budapest, 1121, Hungary}
\email{}

\author[0000-0001-6876-8284]{P. A. Mazzali}
\affiliation{Astrophysics Research Institute, Liverpool John Moores University, Liverpool Science Park, 146 Brownlow Hill, Liverpool L3 5RF, UK}
\affiliation{Max-Planck-Institut f\"ur Astrophysik, Karl-Schwarzschild Str. 1, D-85741 Garching, Germany}
\email{P.Mazzali@ljmu.ac.uk}

\author[0000-0003-0955-9102]{T. Kravtsov}
\affiliation{Department of Physics and Astronomy, University of Turku, FI-20014 Turku, Finland}
\affiliation{Finnish Centre for Astronomy with ESO (FINCA), Quantum, Vesilinnantie 5, University of Turku, FI-20014 Turku, Finland}
\email{timokrav@gmail.com}

\author[0000-0002-8111-4581]{K. Matilainen}
\affiliation{Department of Physics and Astronomy, University of Turku, FI-20014 Turku, Finland}
\email{katja.matilainen@utu.fi}

\author[0000-0002-1022-6463]{T. M. Reynolds}
\affiliation{Niels Bohr Institute, University of Copenhagen, Jagtvej 128, 2200 Copenhagen N, Denmark}
\affiliation{Tuorla Observatory, Department of Physics and Astronomy, University of Turku, FI-20014 Turku, Finland}
\affiliation{Cosmic Dawn Center (DAWN)}
\email{treynolds1729@gmail.com}

\author[0000-0002-1296-6887]{L. Galbany}
\affiliation{Institute of Space Sciences (ICE, CSIC), Campus UAB, Carrer de Can Magrans, s/n, E-08193 Barcelona, Spain}
\affiliation{Institut d'Estudis Espacials de Catalunya (IEEC), E-08034 Barcelona, Spain}
\email{lluisgalbany@gmail.com}

\author[0000-0002-3231-1167]{Bo Wang} 
\affiliation{International Centre of Supernovae (ICESUN), Yunnan Key Laboratory of Supernova Research, Yunnan Observatories, Chinese Academy of Sciences (CAS), Kunming, 650216, China}
\email{wangbo@ynao.ac.cn}

\begin{abstract}

We present a photometric and spectroscopic study of the superluminous Type Ia supernova SN~2022erq. Its early spectra, dominated by iron-group elements with weak intermediate-mass features, might indicate highly efficient nuclear burning, broadly similar to that inferred for some overluminous Type Ia supernovae. The rapid emergence and persistence of narrow Balmer emission lines superposed on this iron-rich spectrum provide clear evidence of long-lived interaction with a hydrogen-rich circumstellar medium (CSM), establishing SN 2022erq as a member of the rare Ia-CSM class. {SN 2022erq reached a peak bolometric luminosity of $\sim 8 \times 10^{43}\ \mathrm{erg\ s^{-1}}$ and exhibited an exceptionally slow post-peak decline, indicating that its light curve is dominated by long duration ejecta–CSM interaction. By combining H$\alpha$ diagnostics with bolometric light-curve modeling, we reconstruct the preexplosion mass-loss history of the progenitor. The mass-loss rate escalated by one order of magnitude over the final decades, rising from $\sim 0.04$ to $\sim 0.6$ M$_\odot$ yr$^{-1}$. This surge produced a massive, extended CSM shell of $\sim 3$~M$_\odot$ out to $\sim 3.5\times10^{16}$ cm. }
The young stellar environment ($\sim 100$\,Myr)  together with this substantial, extensive CSM points to a progenitor system consisting of a white dwarf and an intermediate-mass companion that underwent increasing mass loss prior to explosion. 

\end{abstract}


\counterwithout{figure}{section}
\section{Introduction}
\label{sec:intro}
Type Ia supernovae (SNe Ia) are fundamental tools for measuring cosmic distances and the expansion history of the Universe \citep{1998AJ....116.1009R, 1999ApJ...517..565P}. However, the nature of their progenitor systems remains unresolved. The two leading scenarios are the single-degenerate (SD) model, in which a carbon–oxygen white dwarf (WD) accretes matter from a nondegenerate companion until it approaches the Chandrasekhar mass ($M_{\rm Ch}\approx1.4$\,\Msun)  and undergoes a thermonuclear explosion \citep{2004MNRAS.350.1301H, Wang2018RAA}, and the double-degenerate (DD) model, involving the merger of two WDs \citep{2013ApJ...770L...8P}.

Observational clues to distinguish these channels are often ambiguous. While the general absence of circumstellar material (CSM) around most SNe Ia favours the DD scenario \citep{2007ApJ...670.1275L, 2012ApJ...744L..17B}, the detection of time-variable \NaI\ absorption lines in some events points to the presence of nonuniform CSM, consistent with residual material from an SD companion \citep{2007Sci...317..924P, 2011Sci...333..856S}. 

More direct evidence for preexplosion mass loss comes from SNe Ia that exhibit clear signatures of CSM interaction. SN 2002ic was the first case, displaying narrow H$\alpha$ emission characteristic of shocked, dense CSM and establishing Ia-CSM subclass \citep{2003Natur.424..651H, 2004AstL...30...65C, 2004MNRAS.355..627C, 2004ApJ...605L..37D, 2004ApJ...604L..53W}. The core-degenerate  (CD) scenario, involving a WD merging with the core of an asymptotic giant branch (AGB)  star, has also been proposed for such events \citep{2011MNRAS.417.1466K, 2017MNRAS.464.3965W}.  Subsequent discoveries, including SNe 2005gj \citep{2006ApJ...650..510A, 2007arXiv0706.4088P}, 2018evt \citep{2023MNRAS.519.1618Y, 2024NatAs...8..504W}, and 2020uem \citep{2023ApJ...944..203U, 2023ApJ...944..204U}, have further solidified this class, challenging pure DD models and underscoring the role of enriched circumstellar environments.

This persistent progenitor ambiguity complicates the use of SNe Ia as precision cosmological tools, as different explosion channels could systematically affect their luminosity calibration and, consequently, inferences about dark energy \citep{2011NatCo...2..350H}. Discrepancies with binary population synthesis models add to the challenge of reconciling predicted and observed SN Ia rates \citep{2014ARA&A..52..107M}.

A key manifestation of SN Ia diversity is the wide range of observed peak luminosities. At the bright end of the Phillips relation \citep{1993ApJ...413L.105P}, SN 1991T-like events (e.g., \citealt{1992ApJ...384L..15F, 1992AJ....103.1632P}) are characterized by strong iron-group element (IGE) features and weak intermediate-mass element (IME) lines. Their high nickel yields (e.g., $\sim 1$\,\Msun\ for SNe 1991T and 2011hr; \citealt{2014MNRAS.445..711S, 2016ApJ...817..114Z}) likely result from efficient burning, yet remain consistent with the standard Chandrasekhar-mass regime \citep[e.g.,][]{2022ApJ...930...70H, 2024ApJS..273...16P}. A separate group of even more luminous ``super-Chandrasekhar'' (SC) candidates, such as SNe 2003fg, 2006gz, 2007if, and 2009dc, challenge this limit with inferred ejecta or nickel masses potentially exceeding $M_{\rm Ch}$ \citep{2006Natur.443..308H, 2007A&A...465L..17H, 2007ApJ...669L..17H, 2009ApJ...690.1745M, 2010ApJ...713.1073S, 2010ApJ...715.1338Y, 2009ApJ...707L.118Y, 2010ApJ...714.1209T, 2011MNRAS.410..585S, 2011MNRAS.412.2735T}. Proposed channels for these extreme events include rapidly rotating SD progenitors \citep{2014MNRAS.445.2340W} and massive DD mergers \citep{2018MNRAS.473.5352L}.

\begin{figure*}[!th]
 \centering
 \includegraphics[width=16cm,angle=0]{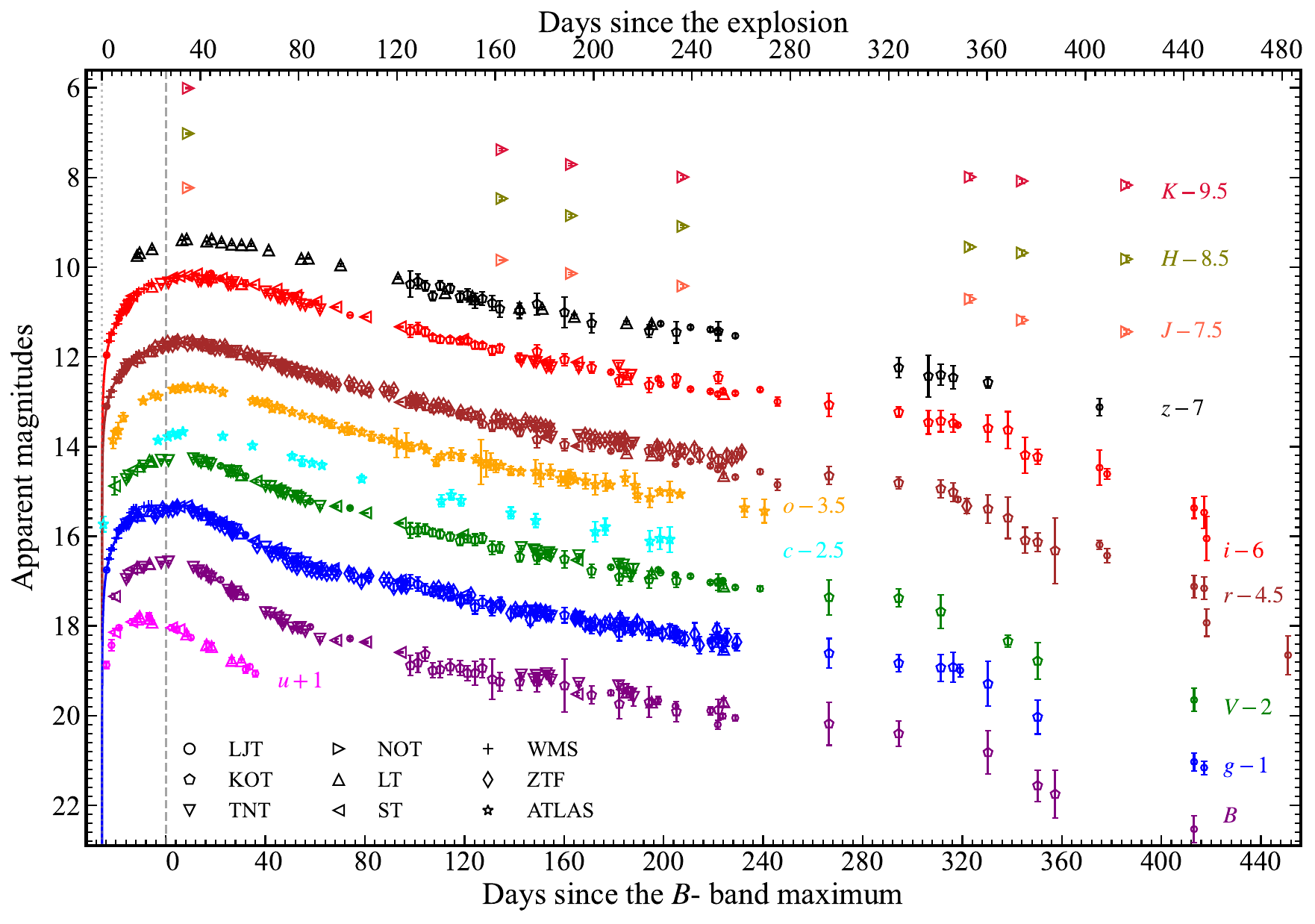}
 \caption{Optical and NIR light curves of SN 2022erq. Dotted and dashed vertical lines mark the explosion epoch and the time of $B$- band maximum, respectively. The data presented in this work are supplemented with public photometry from ZTF and ATLAS. The early-time rise in the $gri$- bands is fitted with a power-law (fireball) model. }
 \label{<lc>}
\end{figure*}

CSM interaction provides an additional pathway to extreme luminosity. In SNe Ia-CSM, the conversion of shock kinetic energy into radiation can augment or even dominate the power source. Some events with early iron-rich spectra plus strong interaction reach luminosities rivaling those of SC candidates \citep{2013ApJS..207....3S,2023ApJ...948...52S}. Such objects are key to mapping progenitor channels, power sources, and potential cosmological systematics.

In this work, we present SN 2022erq, a superluminous ($M_r\approx-21$\,mag) thermonuclear transient that exemplifies this regime. {Its photospheric spectra are dominated by IGE lines accompanied by persistent narrow H$\alpha$ emission from ongoing CSM interaction. Using H$\alpha$ diagnostics and bolometric light-curve modeling, we characterize the CSM properties and reconstruct the preexplosion mass-loss history. We place SN~2022erq within the landscape of superluminous thermonuclear transients and argue that the observed diversity among such events is primarily governed by the presence and properties of CSM, which can dominate the observed energetics even when the underlying explosion produces a substantial mass of IGEs.}

The paper is organized as follows. Section~\ref{obs} describes the observations of SN 2022erq. {Section\,\ref{lc} presents the multiband photometry.} Section \ref{sp} presents the spectral evolution and diagnostics. Section~\ref{dis} discusses the environment, {bolometric luminosity}, CSM properties, and progenitor system. Section~\ref{summ} summarizes the main conclusions.

\begin{figure}
\centering
\includegraphics[width=8.3cm,angle=0]{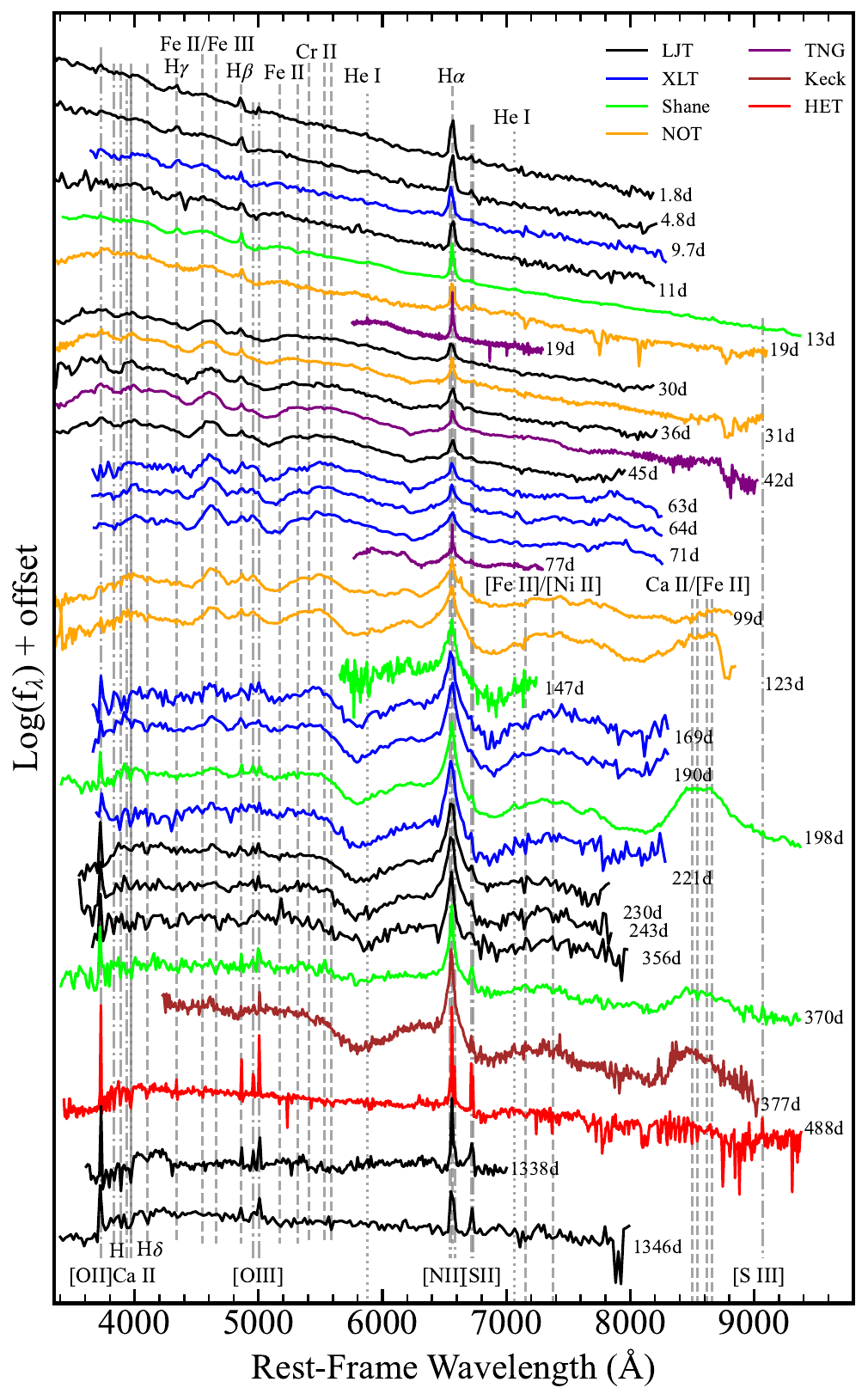}
\caption{Spectral sequence of SN 2022erq. {Epochs marked on the right side of each spectrum are relative to the adopted explosion date.} Dashed and dotted lines mark the rest-frame wavelengths of features originating from the SN and the host galaxy, with identifications labeled above and below the spectrum, respectively. {All spectra have been corrected for the host redshift and  smoothed with bin sizes chosen according to the original signal-to-noise ratio. The data behind this figure are available in machine-readable format in the online journal.}}
\label{<sp>}
\end{figure}

\section{Observations}
\label{obs}
\subsection{Discovery and Classification}
\label{sec:discovery}
SN 2022erq was discovered by the Asteroid Terrestrial-impact Last Alert System (ATLAS; \citealp{2022TNSTR.679....1T}) on 2022 March 11.59 (UTC dates are used throughout this work). A spectrum obtained on 2022 March 12.90 revealed narrow Balmer emission lines superposed on a continuum with broad iron-group absorption features, leading to its classification as an SN 2005gj–like event \citep{2022TNSCR.701....1L}.

This SN is located at R.A.\,=$\,18^{\rm hr}\,33^{\rm m}\,25^{\rm s}$.36, Dec.\,=\,$+44^\circ$\,05\arcmin\,11.65\arcsec  (J2000), residing in a faint irregular dwarf galaxy. A redshift $z\,=\,0.0653\pm0.0001$ is measured from narrow host-galaxy emission lines (e.g., [\OIII] $\lambda\lambda$4959, 5007; [\SII] $\lambda\lambda$6716, 6731) present in late-time spectra (Figure \ref{<sp>}). Adopting H$_0 = 73.04$\,km\,s$^{-1}$\,Mpc$^{-1}$ \citep{2022ApJ...934L...7R}, the corresponding distance is $D = 270 \pm 30$\,Mpc. 

\subsection{{Photometry}}
\label{sec:photometry}
SN 2022erq was monitored extensively in the $BVugrizJHK$ bands using a network of facilities (see Appendix \ref{app:Faci}). For the optical $BVugriz$- bands, host-galaxy subtraction was performed using template images. Specifically, the $griz$ templates were constructed from preexplosion Pan-STARRS archival data. For the $u$ and $BV$ bands, template images were obtained in 2025 November, when the SN had faded sufficiently and its spectrum showed virtually no remaining interaction signatures. Since the $u$ and $BV$ bands do not cover the H$\alpha$ emission line, any residual CSM interaction contribution in these templates is negligible. Point-spread-function (PSF) photometry was applied to the near-infrared (NIR) $JHK$- band data.

Figure~\ref{<img>} provides a finder chart for SN~2022erq and the local reference stars used for photometric calibration. The standardized magnitudes of these reference stars are listed in Table~\ref{refStar}. They were calibrated using photometry from the Pan-STARRS catalog ($griz$ bands; \citealp{2020ApJS..251....6M}), the 2MASS All-Sky Survey ($JHK$ bands; \citealp{2006AJ....131.1163S}), and observations of photometric standard stars ($BV$ and $u$ bands). Transformations from instrumental to standard magnitudes for the $BV$ and $u$ bands were derived on photometric nights using standard stars, e.g., Landolt standards for the $BV$ bands \citep{1992AJ....104..340L} and SDSS standard stars for the $u$ band. For each filter, we solved for the photometric zero points and colour terms, and applied these transformations to the local sequence stars in the SN~2022erq field. The SN magnitudes were then obtained by differential photometry relative to these calibrated local standards and, where necessary, fine-tuned by enforcing consistency between repeated measurements and between overlapping datasets from different instruments. All instrumental magnitudes of SN~2022erq were calibrated against these local reference stars. The final, calibrated light curves are presented in Figure~\ref{<lc>}, with the complete photometric dataset provided in Tables~\ref{tab:photo} and \ref{app_a:JHK}. For comparison, we also plot publicly available Zwicky Transient Facility (ZTF\footnote{\url{https://alerce.online/object/ZTF22aadgsdi}}; \citealp{2019PASP..131a8002B,2019PASP..131g8001G}) $gr$-band data and ATLAS\footnote{\url{https://fallingstar-data.com/forcedphot}}(\citealp{2018PASP..130f4505T,2020PASP..132h5002S}) $c$- and $o$-band measurements obtained from forced photometry in Figure~\ref{<lc>}.

The last ATLAS non-detection in the $c$ band occurred on 2022 March 9.66. Together with the subsequent ATLAS discovery on March 11.59 in the same band, this provides an initial constraint on the explosion epoch of MJD = 59649.6 $\pm$ 1.0. We further refine the explosion epoch to MJD = 59649.1 $\pm$ 0.5 by fitting the early-time $gri$- band light-curve rise with a simple fireball model (Figure~\ref{<lc>}), which assumes a uniformly expanding photosphere with constant surface luminosity \citep{1982ApJ...253..785A}. This estimate is consistent with the ATLAS non-detection, and we therefore adopt MJD = 59649.1 as the reference explosion time hereafter.

\subsection{Spectroscopy}
With the explosion epoch established, the first spectrum of SN 2022erq was obtained at $\tau\approx 1.8$\,d (where $\tau$ denotes days since explosion and $t$ refers to days after $B$- band maximum), making it the earliest spectroscopically confirmed SN Ia‑CSM event. We conducted a spectroscopic campaign from $\tau\approx 2$\,d to 1350\,d using multiple facilities (Appendix \ref{app:Faci}). The spectral sequence is presented in Figs. \ref{<sp>} and \ref{fig:SpNIR}, with observational details provided in Table \ref{Tab:Spec_log}.  

{All spectra were processed with standard reduction pipelines, which included wavelength calibration, flux calibration, and correction for telluric absorption. The spectra were also shifted to the rest frame using the host-galaxy redshift. To minimize the effects of atmospheric dispersion, most observations were conducted with the spectrograph slit aligned at the parallactic angle \citep{1982PASP...94..715F}. For each spectrum, we then refined the flux scale and continuum shape using multiband photometry obtained at closely matching epochs.}

For the subsequent analysis, all spectra were corrected for line-of-sight extinction. The absence of host-galaxy \NaI\ absorption indicates negligible host reddening, consistent with our environmental modeling (Section~\ref{Gal}) which yields $A_V({\rm host})\approx0.046$\,mag and $E(B\!-\!V)_{\rm host}\approx0.015$\,mag (assuming $R_V=3.1$). Including Milky Way reddening, $E(B\!-\!V)_{\rm MW}=0.0046\pm0.0005$~mag \citep{2011ApJ...737..103S}, we adopt a total $E(B\!-\!V) =0.02$\,mag.

\section{{Photometric Analysis}}
\label{lc}
\subsection{K-Correction and Measurement}
The observed $BVugriz$- band light curves of SN~2022erq were corrected for redshift-dependent effects using the K-correction method described in \citet{2007ApJ...663.1187H}. For each photometric epoch, we applied a K-correction derived from a spectrum of SN 2022erq obtained at a similar phase. The resulting K-corrections as a function of time for each filter are shown in Figure \ref{<Kcor>}. All subsequent light-curve analyses are based on these K-corrected data.

We measured light-curve parameters by fitting the K-corrected data with low-order polynomials.  The resulting quantities, summarized in Table~\ref{Tab:pho_results}, include the peak time ($t_{\rm max}$), rise time ($t_{\rm rise}$), peak magnitude ($m_{\rm peak}$), absolute magnitude ($M_{\rm peak}$), and post-maximum decline of the light curve in 15 d ($\Delta m_{15}$). 

Consistent with other Ia-CSM events, SN~2022erq shows a clear increase in the rise time from the blue to the red bands \citep{2007arXiv0706.4088P,2023ApJ...948...52S}. It reaches maximum light in the $u$ band at $\tau \approx 17.5$~d, whereas the $z$-band peak occurs much later, at $\tau \approx 40.7$~d. This wavelength-dependent delay partially reflects the temperature evolution of the spectral energy distribution (SED), a common feature among most SNe. In SN~2022erq, however, the particularly slow evolution in the $r$ band is further enhanced by the gradual strengthening of H$\alpha$ emission, which contributes additional flux at this wavelength and prolongs the rise toward maximum.

The $JHK$-band light curves are more sparsely sampled. The earliest NIR epoch at $\tau \approx 34.6$ d likely precedes the true maxima, so the NIR peak magnitudes, estimated from the earliest data near maximum, should be regarded as lower limits: $M^J_{\rm peak} \approx -21.45$, $M^H_{\rm peak} \approx -21.65$, and $M^K_{\rm peak} \approx -21.66$ mag.

\begin{table}
\centering
\tabletypesize{\scriptsize}
\caption{Light curve parameters of SN 2022erq}
\begin{tabular}{cccccc}
\hline\hline
Band   &$t_{max}$      & $t_{rise}$  &  $m_{peak}$  & $M_{peak}$ & $\Delta m_{15}$      \\
       &(MJD)          &  (d)        &  (mag)       & (mag)      & (mag)   \\
\hline
$u$	   	&	59667.76	&	17.51	&	16.72	&	-20.51	&	0.38	\\
$B$	   	&	59674.70	&	24.03	&	16.55	&	-20.67	&	0.23	\\
$g$	   	&	59678.89	&	27.96	&	16.33	&	-20.89	&	0.22	\\
$V$    	&  	59679.52	&  	28.56	&  	16.30	&	-20.90	&	0.19	\\
$r$	   	&  	59681.57	&  	30.48	&  	16.15	&	-21.05	&	0.15	\\
$i$	   	&  	59687.64	&  	36.18	&  	16.21	&	-20.98	&	0.11	\\
$z$	   	&	59692.50	&	40.74	&	16.33	&	-20.85	&	0.10	\\
\hline
\hline
\end{tabular}

\raggedright
\tablecomments{The rise time and the decline rates are calculated in the rest frame (corrected for the time-delay effect due to cosmological redshift). Typical uncertainties are $<$ 1 d for time measurements, 0.01–0.02 mag for apparent magnitudes and decline rates, and  $\sim$ 0.1 mag for absolute magnitudes, dominated by the distance uncertainty. }
\label{Tab:pho_results}
\end{table}

\begin{table}
\centering
\scriptsize
\caption{$B$-band light-curve parameters}
\begin{tabular}{lcccccc}
\hline\hline
SN &  $t_{\rm rise}$ &$M_{\rm max}$ &$\Delta m_{15}$ & $\Delta m_{50}$ & $z$ & Ref.    \\
    & (d)       & (mag)     & (mag)    & (mag)     &    & \\
\hline
2022erq &	24.03	&	-20.67	&	0.23	&	1.42	&	0.0653	&	(1)	\\
2011fe &	17.26	&	-19.24	&	1.18	&	3.36	&	0.0012	&	(2)	\\
1991T & 	17.90	&	-19.98	&	0.95	&	3.04	&	0.0058	&	(3)	\\
2011hr &	19.17	&	-19.84	&	0.93	&	3.03	&	0.0133	&	(4)	\\
2007if &	20.67	&	-20.53	&	0.76	&	2.77	&	0.0742	&	(5)	\\
2009dc &	21.83	&	-20.22	&	0.73	&	2.78	&	0.0216	&	(6)	\\
\hline\hline
\end{tabular}

\raggedright
\tablecomments{$B$-band Light curve parameters of SN 2022erq with those of  normal SN~Ia SN 2011fe, 91T-like (SN 1991T, SN 2011hr) and SC candidates (SN 2007if, SN 2009dc). The rise times and the decline rates are calculated in the rest frame (corrected for the time-delay effect due to cosmological redshift). References: (1) This work, (2) \cite{2016ApJ...820...67Z}, (3) \cite{2014MNRAS.445..711S}, (4) \cite{2016ApJ...817..114Z}, (5) \cite{2010ApJ...713.1073S}, (6) \cite{2010ApJ...714.1209T}}

\label{Tab:LCpare}
\end{table}

\begin{figure*}
\centering
\includegraphics[width=15cm,angle=0]{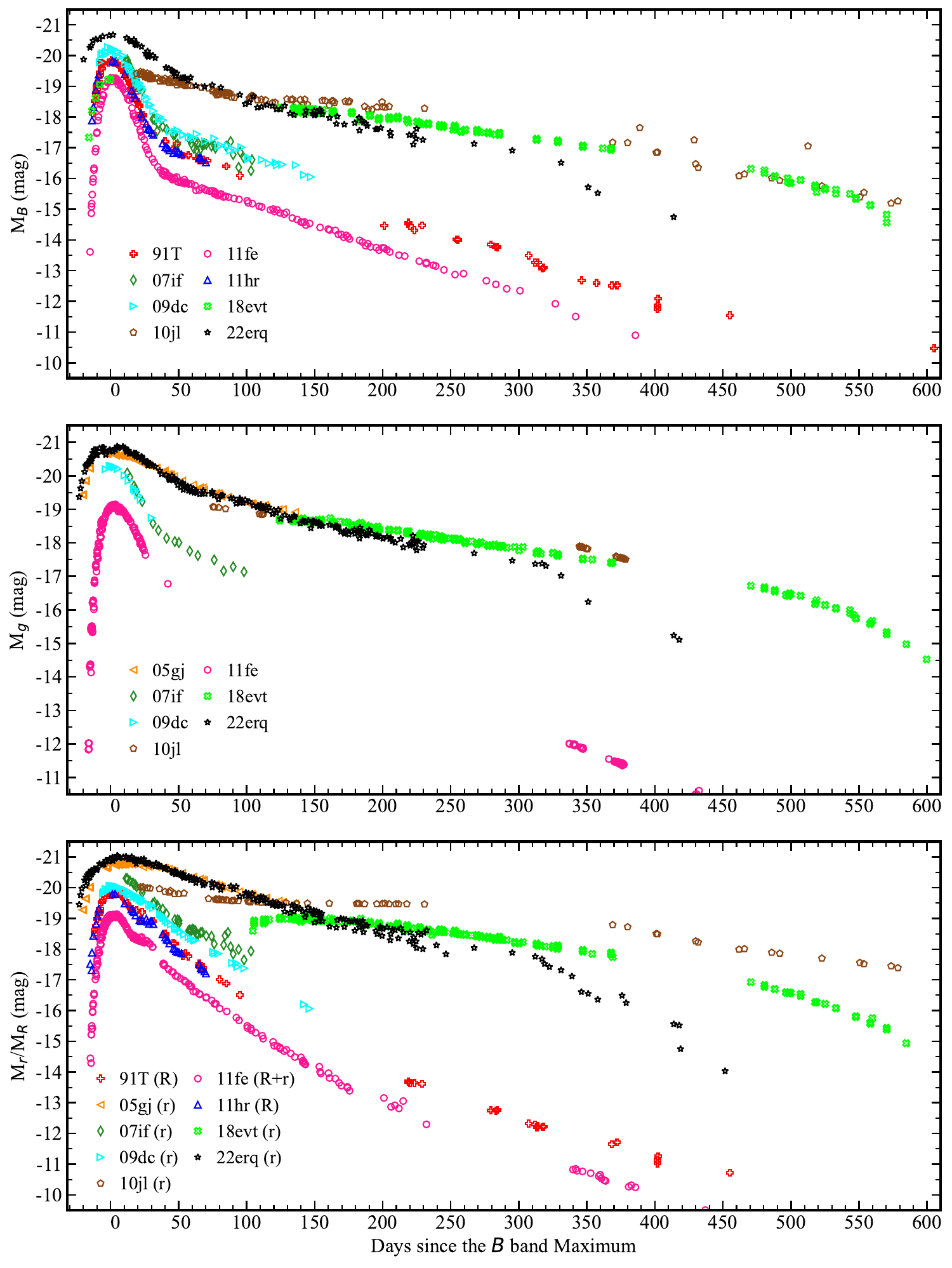}
\caption{The $B$-, $g$-, and $r/R$-band light curves of SN 2022erq compared with representative events: SNe Ia-CSM (SNe 2005gj, 2018evt; \citealp{2006ApJ...650..510A,2007arXiv0706.4088P,2023MNRAS.519.1618Y,2024NatAs...8..504W}), SC candidates (SNe 2007if, 2009dc; \citealp{2010ApJ...713.1073S,2011MNRAS.412.2735T,2012ApJS..200...12H,2012MNRAS.425.1789S,2019MNRAS.490.3882S}), 91T-like SNe (SNe 1991T, 2011hr; \citealp{1992ApJ...384L..15F,1992AJ....103.1632P,1994ApJ...434L..19S,2016ApJ...817..114Z}), the normal SN Ia SN 2011fe \citep{2011Natur.480..344N,2013NewA...20...30M,2015MNRAS.446.3895F,2016ApJ...820...67Z}, and the Type IIn SN~2010jl \citep{2011ApJ...730...34S,2012AJ....143...17S,2014ApJ...789..104O,2014ApJ...797..118F}. Where $r$- band coverage is insufficient or unavailable, $R$- band data are used as a complement or substitute. }
\label{BrLC}
\end{figure*}

\subsection{Light Curve}
The photometric parameters of SN 2022erq are summarized in Table \ref{Tab:pho_results}. A comparison with normal SNe Ia, luminous 1991T-like events, and SC candidates (Table~\ref{Tab:LCpare}) demonstrates its extreme luminosity relative to these classes. At peak, SN 2022erq is $\sim$ 1.5 mag brighter in the $B$- band than a typical SN Ia. This excess increases to $\sim$ 2.5~mag by $t = +15$ d and to $\sim$ 3.5 mag by $t = +50$~d. The contrast is even stronger in the $r$- band, consistent with enhanced longer-wavelength emission from the shocked CSM, e.g., H$\alpha$  emission, \citep{2023ApJ...948...52S}.

Figure \ref{BrLC} compares the multiband light curves of SN 2022erq with those of a broader sample, extending the comparison in Table \ref{Tab:LCpare} to include the Ia-CSM SN~2005gj and SN 2018evt, and the Type IIn SN 2010jl. SN~2022erq is slightly more luminous than SN 2005gj, but their light curves share a similar morphology. Both exhibit a slow rise and a gradual decline, characteristic signature of early  sustained CSM interaction that governs their photometric evolution. The CSM interaction in SN 2022erq appears to weaken at $t \gtrsim +330$~d, as indicated by a clear steepening of the decline rate in all bands.

\begin{figure}
\centering
\includegraphics[width=8.3cm,angle=0]{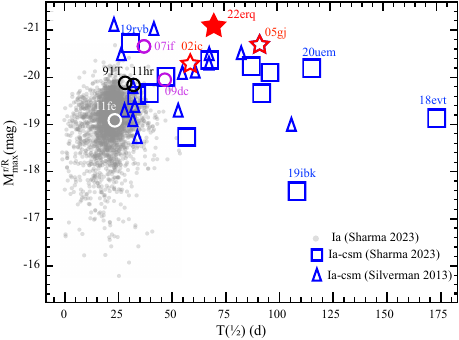}
\caption{Peak $r/R$-band luminosity ($M^{r/R}_{\rm max}$) vs. duration above half-maximum luminosity ($T_{1/2}$) in the rest frame. The sample includes normal, 91T-like, SC candidate, and CSM-interaction SNe Ia from \cite{2013ApJS..207....3S,2023ApJ...948...52S}.}
\label{<LCcomp>}
\end{figure}

SN 2018evt, in contrast, was fainter at early times, consistent with its weaker early-time spectroscopic signatures of interaction.  Although its late-time light curve indicates prolonged interaction, the lack of very early observations precludes a precise constraint on the onset of its CSM-interaction phase. 
When  re-observed from $t \gtrsim +$120 d onward, SN 2018evt exhibited a luminosity comparable to SN 2022erq at $B$- and $g$- bands, and surpassed it after $t\gtrsim +200$ d.  It evolved more slowly, with a sharp downturn after $t\gtrsim +$ 500~d. Given the  spectral similarity of these two SNe at  $\tau \sim 150-250$~d (Section \ref{comp}), which implies comparable ejecta velocities, the higher late-time luminosity and the delayed rapid decline of SN 2018evt indicate that its outer CSM is likely both denser (or more optically thick) and more spatially extended than that of SN 2022erq.

Figure \ref{<LCcomp>} places SN 2022erq in a broader context by plotting peak $r/R$- band absolute magnitude against light-curve width (duration above half-maximum) for SNe Ia, including Ia‑CSM events. SN 2022erq lies at the luminous extreme of this parameter space. The diversity in peak luminosity and light-curve evolution among these interacting transients results from differences in the onset strength of the SN–CSM interaction and in the efficiency of converting shock kinetic energy to radiation. SN 2022erq represents a case of exceptionally early and efficient coupling.

\begin{figure}
\centering
\includegraphics[width=8.2cm,angle=0]{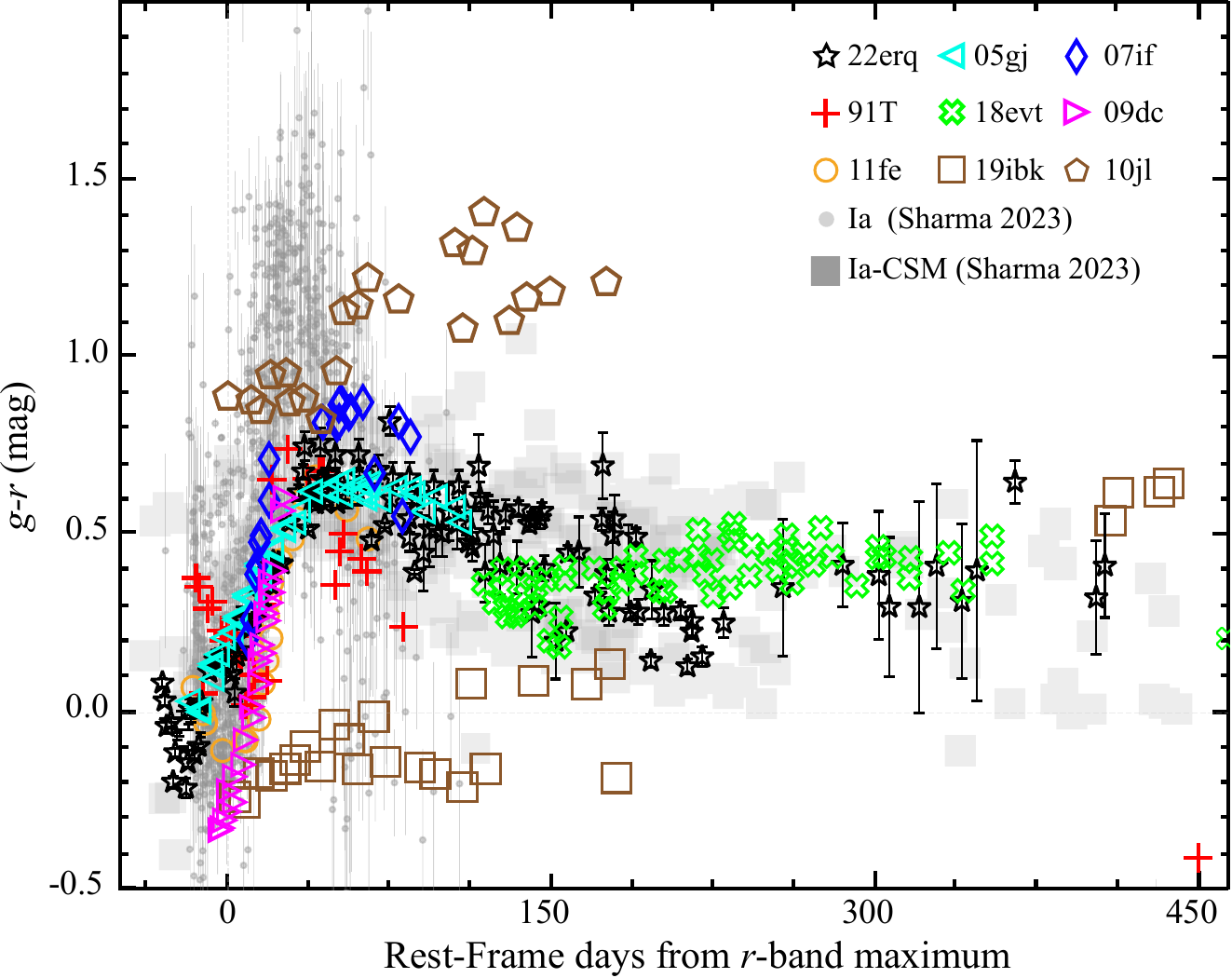}
\caption{The $g-r$ color evolution of SN 2022erq compared to representative objects, including a sample of SNe Ia and Ia-CSM from \citet{2023ApJ...948...52S}, as well as several other well-studied transients. All comparison objects have been corrected for reddening. }
\label{<cc>}
\end{figure}

\subsection{Color Curve}
Figure~\ref{<cc>} compares the $g-r$ color evolution of SN~2022erq with a sample of SNe~Ia, including SNe~Ia-CSM, and other well-studied transients. SN~2022erq closely tracks the evolution of luminous SNe~Ia-CSM like SN~2005gj. With the very early coverage, SN 2022erq shows an initial blueward drift as the temperature rises, reaching the bluest $g-r$ color roughly 2~weeks before the $r$- band maximum, followed by a gradual reddening, which is a pattern broadly analogous to normal SNe~Ia.

Around the maximum light, SN 2022erq is redder than normal SNe Ia. Its $V$-band peak is $\sim\,0.2$ mag brighter than its $B$-band peak, whereas in normal SNe Ia these two bands typically peak at a similar brightness \citep{2009ApJ...697..380W}. The relative redness is consistent with the finding that SNe Ia-CSM are generally cooler than normal SNe Ia at early times \citep{2023ApJ...948...52S}.

SN~2022erq reaches its reddest $g-r$ color at $\sim 50$ d after the $r$- band maximum, with a value of $\sim 0.7$ mag smaller than typical SNe Ia.  The luminous SN 1991T and the SC candidate SN 2007if reach a comparable maximum $g-r$ colors to SN 2022erq.  

Thereafter, the $g-r$ color of SN 2022erq and other Ia-CSM SNe evolves slowly, settling onto a plateau of $\sim 0.5$ mag. This sustained plateau coincides in time with the formation and persistence of a cold dense shell (CDS) seen in the spectra (Section \ref{csm}), and is likely maintained by ongoing CSM interaction (continuous energy input and possible obscuration). In contrast, the $g-r$ color of normal SNe Ia, SN 1991T, and SN 2007if declines rapidly after reaching its maximum,  reflecting the absence of the sustained CSM-interaction phase that dominates the late-time emission.

The similarity in the color evolution between luminous SNe~Ia-CSM like SN~2005gj and SN~2022erq may point to a common physical origin. However, not all SNe~Ia-CSM follow this behavior. 
The subluminous Ia-CSM SN~2019ibk ($M^{r}_{\rm max} \approx -17.5$~mag; \citealp{2023ApJ...948...52S}) remains significantly bluer within $\sim 120$~d after $r$-band maximum, hinting at different progenitor, explosion mechanism, CSM properties or viewing angles. Besides, the Tpe IIn SN 2010jl is even redder than SNe~Ia-CSM, likely due to stronger obscuration of the underlying supernova by a denser, hydrogen-rich CSM.

\section{Spectral Analysis}
\label{sp}

\subsection{Evolution}
\label{Spev}

The long-term spectroscopic campaign of SN 2022erq offers a rare opportunity to trace the evolution of an SN Ia-CSM. Its early-time spectra are dominated by a narrow H$\alpha$ emission. {The origin of this feature, whether from the underlying host galaxy or from SN–CSM interaction, can be reliably diagnosed from its temporal evolution. Owing to the limited spectral resolution of our early data (instrumental Full Width at Half Maximum, FWHM $\approx 700-1000$ \kms), any truly narrow galactic component and a somewhat broader interaction-powered component would be blended, so the line width alone cannot be used as a robust discriminator. Instead, we rely on the evolution of the relative strength of the emission lines.} The significantly weaker relative intensities of galactic emission lines (e.g., [\SII], [\NII], [\OII], [\OIII]) with respect to H$\alpha$ in the early spectra compared to the late-time spectra at $\tau\gtrsim 1300$ d when the SN has faded below the detection threshold, indicate that the initial H$\alpha$ and H$\beta$ emission originate from SN-CSM interaction rather than galactic contamination. 

\begin{figure}
 \centering
 \includegraphics[width=8.3cm,angle=0]{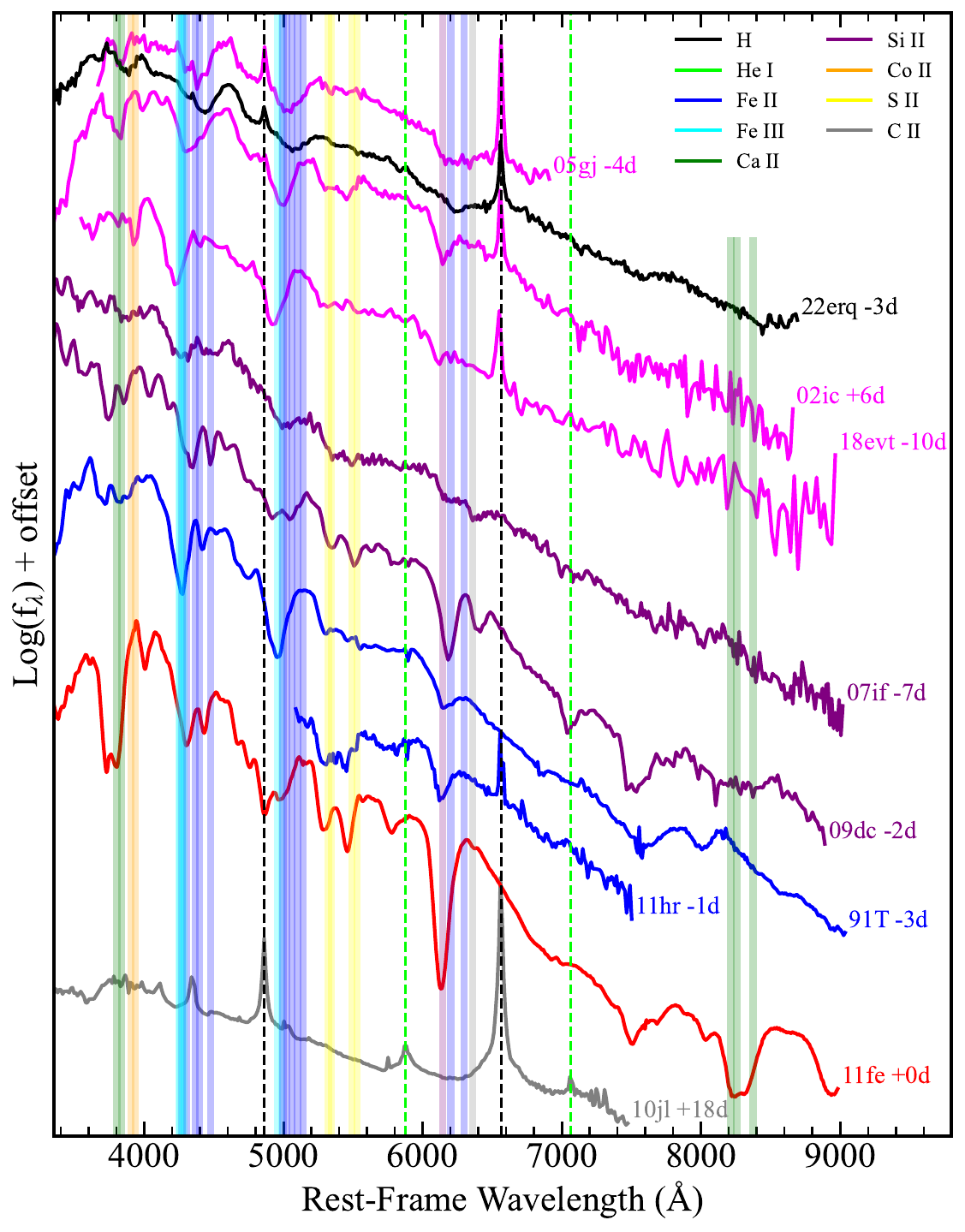}
 \caption{Spectral comparison of SN 2022erq near maximum brightness with SNe~Ia-CSM (SNe 2002ic, 2005gj, 2018evt), SC candidates (SNe 2007if, 2009dc), 91T-like (SNe 1991T, 2011hr), normal SN~Ia SN 2011fe \citep{2012ApJ...752L..26P}, and SN~IIn SN 2010jl \citep{2012AJ....143...17S}. Thin dotted lines mark rest-frame wavelengths; thick lines indicate features at a blueshift of 10,000 \kms. All spectra are dereddened. } 
 \label{fig:Sp_max}
\end{figure}

Absorption from IMEs (e.g., Si, S, Ca) is nearly undetectable until near the time of the $B$-band maximum brightness (Figure \ref{fig:Sp_max}). An emission bump at $\sim 4500$ \AA\ attributed to \FeII\,and \FeIII\,emerges at $\tau\approx 20$ d, persists with stable strength until $\tau \approx 123$ d, and subsequently fades by $\tau\approx 170$ d. Additionally, a similar IGE-produced emission bump {near 5500 \AA, attributed to a blend of permitted lines of \CrII\ (e.g., $\lambda\lambda$ 5410, 5535, 5587) and 
\FeII\ $\lambda$ 5316 \citep{2015MNRAS.448.2766B, 2018MNRAS.474.3931B} } becomes noticeable at $\tau\approx 60$ d, also fading by $\tau\approx 170$ d. Concurrently, the H$\alpha$ line profile evolves slowly from $\tau\approx 100$ d to 377 d.

At $\tau\approx 123$\,d, two broad emission bumps appear near 7300\,\AA\ and 8500\,\AA. The former likely blends [\FeII] $\lambda$7155, [\NiII] $\lambda$7378, and [\FeII] $\lambda$ 7720, while the latter may consist of the \CaII\ NIR triplet  and [\FeII] $\lambda$8617. These features intensify and then weaken, remaining detectable at $\tau\approx 380$\,d. Characteristic nebular-phase lines of normal SNe~Ia, such as numerous Fe and Co emission lines in the blue, are absent in SN 2022erq, likely owing to CSM obscuration. At $\tau\approx 488$\,d, all previously prominent SN features, including the broad H$\alpha$ emission underlying the  narrow component, have largely faded, and the spectrum becomes dominated by host-galaxy lines.

Overall, the spectroscopic evolution of SN 2022erq is notably slow, plausibly maintained by ongoing CSM interaction that converts shock kinetic energy into radiation and prolongs photospheric heating. This sustains high luminosity and delays spectral transitions, consistent with other interacting SNe \citep[e.g.,][]{2017Natur.551..210A,2017hsn..book..403S,2020MNRAS.498...84Z,2025ApJ...978..163Z}.

\begin{figure*}[!htbp]
 \centering
 \includegraphics[width=17.6cm,angle=0]{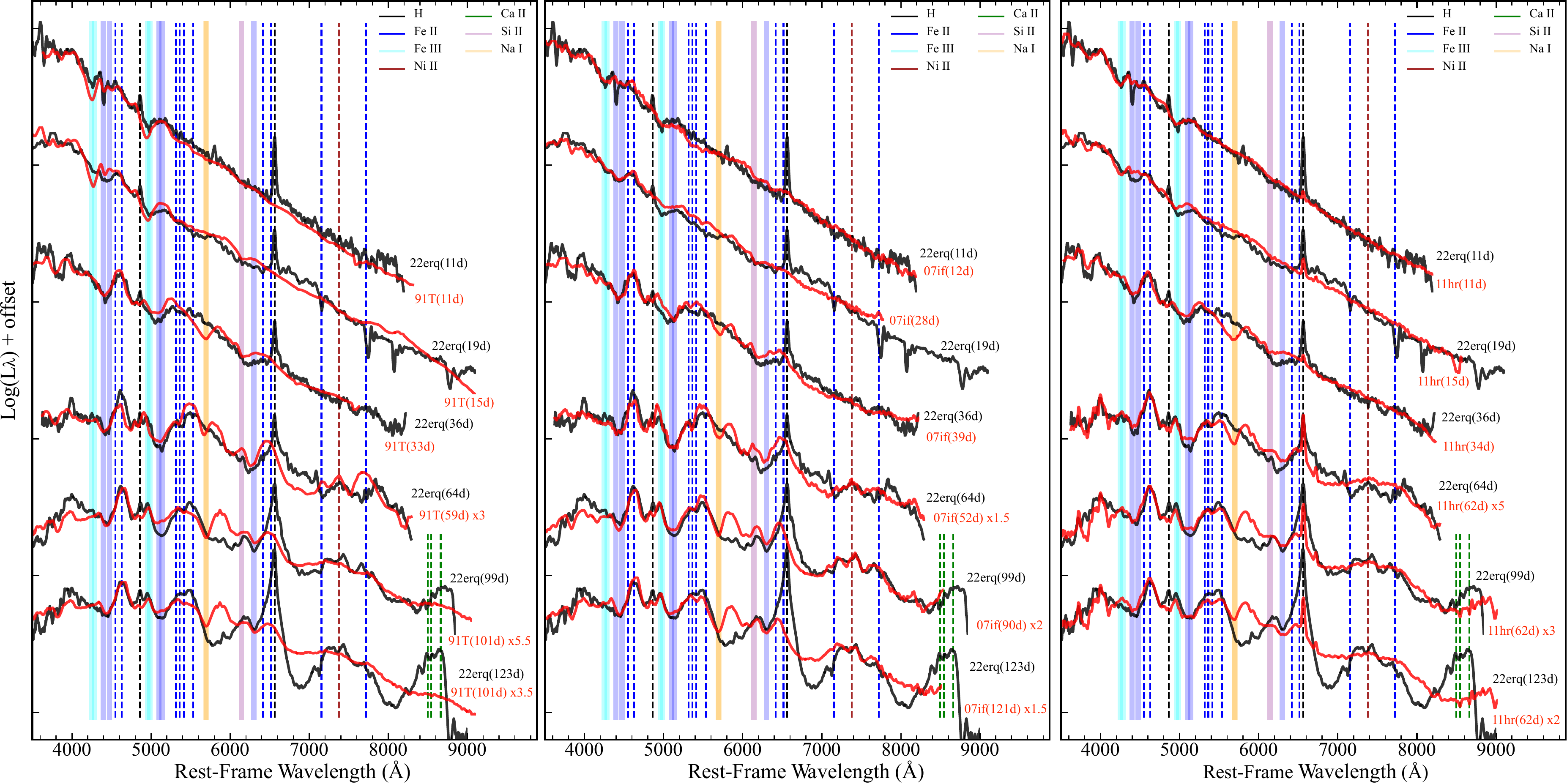}
 \caption{Spectral comparison of SN 2022erq with SNe 1991T, 2007if, and 2011hr.  Luminosity scaling factors, where applied, are noted after the phase.  Thin dotted lines indicate the rest-frame wavelengths of spectral lines, while thick lines show their positions at  a blueshift of 10,000\,\kms.}
 \label{fig:Sp_sub}
\end{figure*}
\begin{figure*}
 \centering
 \includegraphics[width=17.6cm,angle=0]{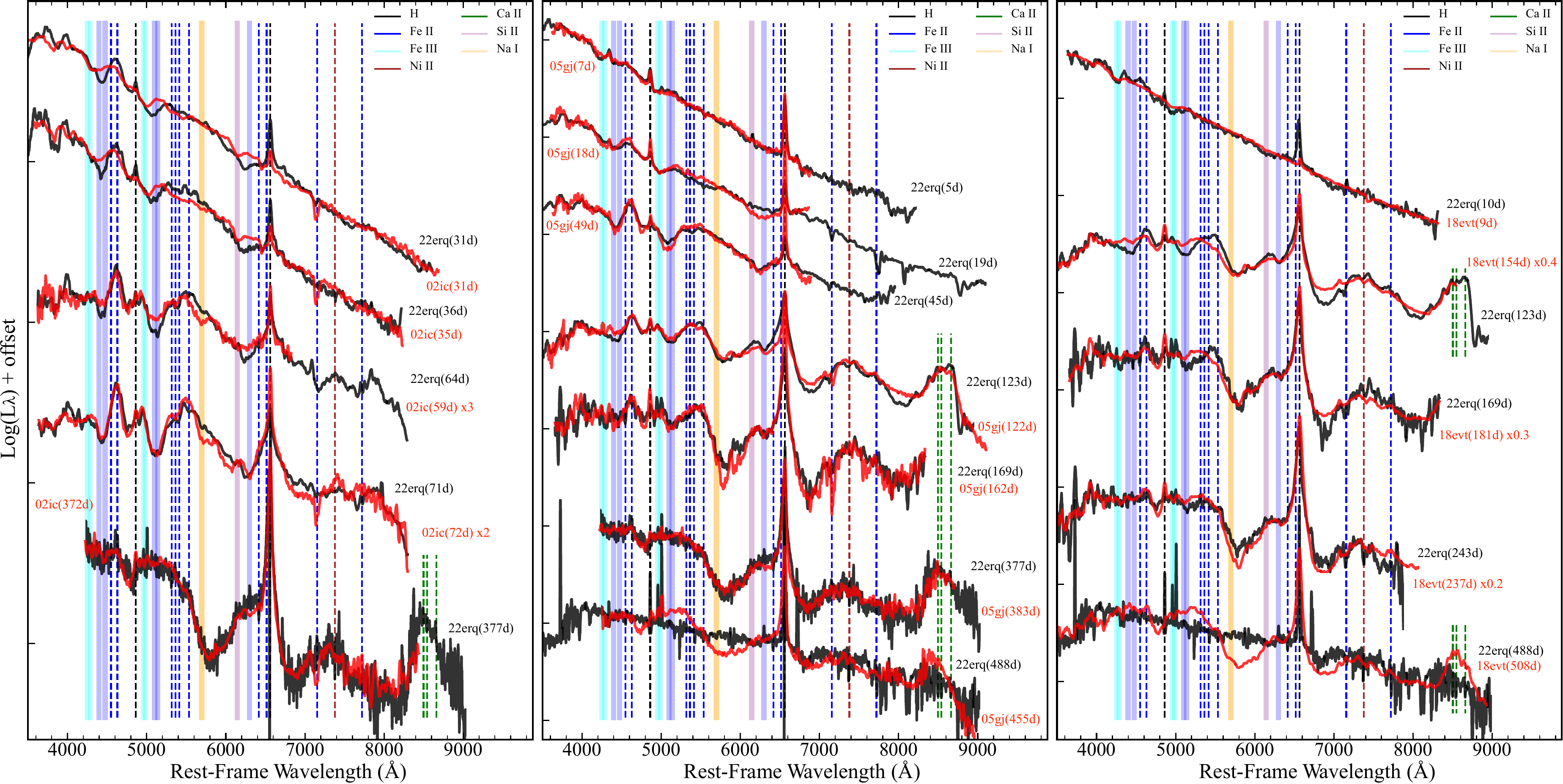}
\caption{SN 2022erq compared with SN~Ia-CSM SNe 2002ic, 2005gj, and 2018evt, following the same matching procedure applied in Figure \ref{fig:Sp_sub}.}
 \label{fig:SpIacsm}
\end{figure*}

\subsection{{Comparison with Iron-Rich SNe Ia}}
\label{comp}
Spectroscopic identification of SN~Ia-CSM events relies on the presence of multicomponent H$\alpha$ emission (relatively narrow core with intermediate/broad base) superimposed on a diluted SN Ia spectrum, first established in SN 2002ic \citep{2003Natur.424..651H}. Subsequent analyses indicated that the underlying SN component more closely resembles luminous, SN 1991T-like SNe Ia with enhanced IGE abundances (e.g., SNe 2005gj and 2018evt; \citealp{2006ApJ...650..510A,2024NatAs...8..504W}).

Figure~\ref{fig:Sp_sub} compares SN~2022erq with spectroscopic templates of transients known for efficient nuclear burning: the luminous SNe~1991T and 2011hr, and the SC candidate SN~2007if (see Appendix~\ref{app:Spmatch} for technical details). The aim of this comparison is to evaluate whether the photospheric signatures of SN~2022erq are consistent with an explosion that produced a high yield of IGEs. This task is challenging because strong CSM interaction boosts the continuum flux and can dilute photospheric absorption features.

In the first month, all comparison templates show broad agreement with SN~2022erq. The spectra are dominated by \FeII/\FeIII\ absorption features near $\sim 10{,}000$~\kms, while IME lines such as Si~II~$\lambda 6355$ are exceedingly weak. This stands in stark contrast to normal SNe~Ia such as SN~2011fe (Figure~\ref{fig:Sp_max}) and points to a common nucleosynthetic outcome characterized by efficient burning and a high IGE yield.

At later phases ($\tau \gtrsim 60$~d), detailed differences emerge. The spectrum of SN~2007if provides a close match across multiple epochs: the [\FeII] 5316–5535~\AA\ quartet, the blended [\FeII]+[\NiII] complex near 7000–8000~\AA, and the cleanly resolved lines [\FeII]~$\lambda 7155$, [\NiII]~$\lambda 7378$, and [\FeII]~$\lambda 7720$ at $\tau \approx 90$~d are all well reproduced. SN~1991T at a comparable epoch ($\sim 59$~d) also shows prominent Fe/Ni emission in this region, confirming a spectroscopic kinship; however, these features appear to fade more rapidly in SN~1991T than in SN~2022erq. SN~2011hr at $\sim 62$~d matches SN~2022erq well at early times, but its spectral coverage ends too soon to assess the later evolution.

A definitive assessment of which individual object is the optimal match is limited by two factors. First, CSM interaction can significantly delay the spectroscopic evolution of the underlying SN, complicating direct phase-to-phase comparisons. Second, the available spectral sequences for the templates are incomplete and not perfectly aligned in time with the extensive coverage of SN~2022erq. Therefore, the comparison does not identify a unique genealogical link to a specific SN~Ia subtype. Instead, it robustly establishes that the spectroscopic phenotype of SN~2022erq, marked by weak IMEs and strong, persistent IGE features, is shared with transients known to arise from highly efficient thermonuclear burning. This interpretation is supported by the low IME masses (e.g., $\sim 0.1$–0.2~\Msun) derived from detailed spectral modeling of SNe~1991T and 2011hr \citep{2014MNRAS.445..711S, 2016ApJ...817..114Z}.

We therefore conclude that the underlying explosion of SN~2022erq likely belongs to the same family of efficient-burning thermonuclear events as overluminous (SN 1991T-like) SNe~Ia and some SC candidates (e.g., SN~2007if). Its position within this sequence is defined primarily by its nucleosynthetic yield. The subsequent analysis demonstrates that its extraordinary luminosity and light-curve behavior, however, are dictated not by this yield, but by the conversion of kinetic energy via interaction with its massive, hydrogen-rich CSM. 

\subsection{Comparison with Other CSM‑interacting SNe}
\label{comp2}

A comparison with other SNe Ia-CSM (Figure \ref{fig:SpIacsm}) shows that SN 2022erq most closely resembles SN 2005gj during the first year.  {The nearly identical spectroscopic evolution of SN 2022erq and SN 2005gj, together with the similarity with SN 2007if established above, suggests that SN 2007if may also serve as a useful spectral analog for SN 2005gj, a connection possibly overlooked because SN 2007if was discovered later.} The interaction signatures in SN 2022erq begin to fade after $\tau\approx 380$\,d, a timing that coincides with the clear steepening of the light‑curve decline observed from $t\gtrsim +330$d. This earlier weakening, compared with the prolonged interaction seen in SN 2018evt, implies that the CSM around SN 2022erq is less spatially extended. On the other hand, the stronger early‑time SN Ia signatures (e.g., more distinct IME absorptions)  and the weaker H$\alpha$ emission in SNe 2002ic and 2018evt correlate with their lower peak luminosities, indicating that CSM interaction was weaker at early phases in those events.

SN 2022erq differs fundamentally from SNe IIn in its CSM properties. Its high Balmer decrement (H$\alpha$/H$\beta\approx 7$), approximately double that of SN 2010jl at a comparable phase, aligns with other SNe Ia-CSM \citep{2023ApJ...948...52S} and suggests enhanced continuum opacity or dust extinction. The asymmetric H$\alpha$ profile ($\tau\gtrsim 100$\,d), with a suppressed red wing, may indicate dust formation \citep{2024NatAs...8..504W}. The absence of H$\gamma$ emission, prominent in SN 2010jl (Figure \ref{fig:Sp_max}), might point to a different ionization conditions or excitation states. While no clear reliable narrow optical \HeI\ emission  is seen, a distinct P-Cygni profile in the \HeI\ $\lambda$10,830 line is present in all NIR spectra of SN~2022erq ($13 < \tau < 150$\,d; Figure \ref{fig:SpNIR}), confirming helium in the CSM and suggesting that the optical \HeI\,lines are suppressed due to higher optical depth or different ionization balance.

\begin{figure}
 \centering
 \includegraphics[width=8.3cm,angle=0]{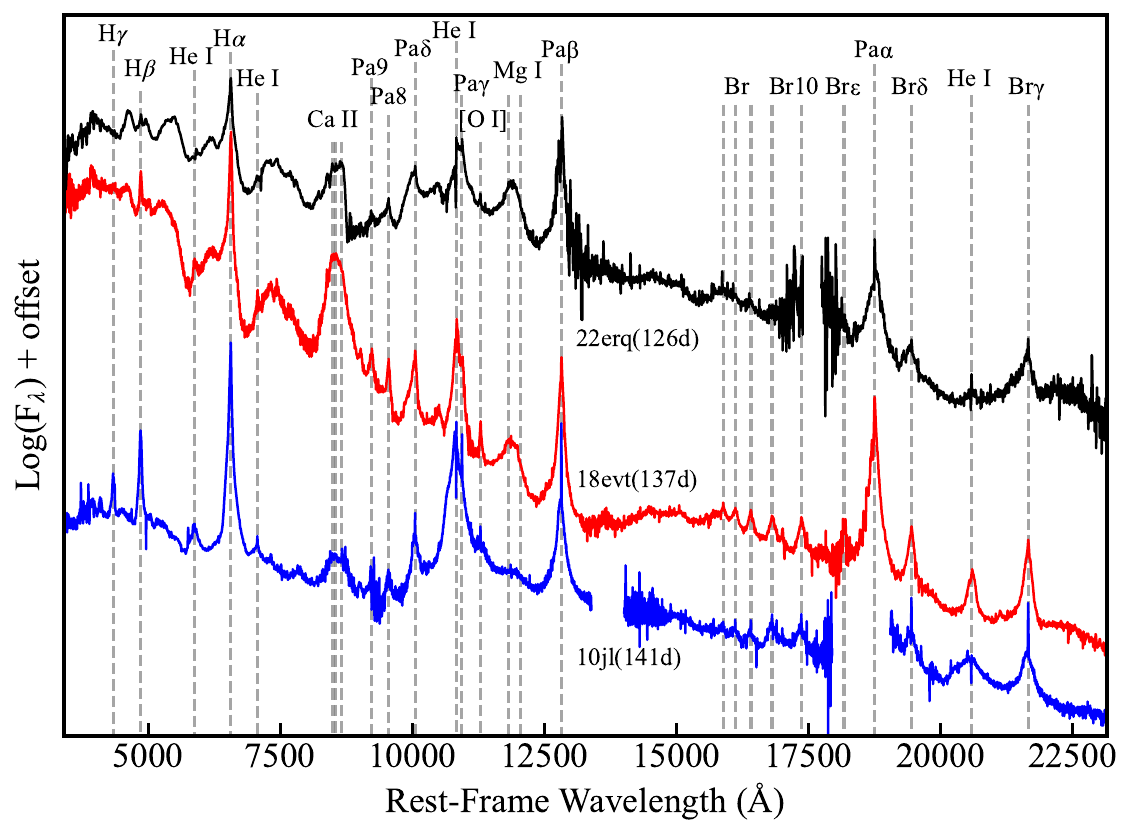}
 \caption{A phase-averaged spectrum of SN 2022erq at $\tau \approx 126$\,d, constructed from adjacent optical and NIR observations, is compared with those of SNe 2018evt and 2010jl \citep{2012AJ....144..131Z,2015ApJ...801....7B}. Dashed lines indicate the rest-frame wavelengths of the spectral lines.} 
 \label{fig:OPNIR}
\end{figure}

The comparison of the optical--NIR spectral properties of SN 2022erq with those of SNe 2018evt and 2010jl (Figure \ref{fig:OPNIR}) highlights both shared and distinct characteristics. All three objects exhibit dominant H emission lines, including the Balmer, Paschen, and Brackett series. The Brackett series lines in SN 2022erq are weaker than in SN 2018evt and SN 2010jl, indicating similar CSM compositions but differing density and temperature.  A spectral bump observed at $\sim 12,000$ \AA\ in SN 2022erq and SN 2018evt may originate from \MgI\,$\lambda\lambda$11,828, 12,047, which is less pronounced in SN 2010jl, suggesting differences in their inner SN compositions.

\section{Discussion}
\label{dis}

\subsection{Galactic Environment}
\label{Gal}

Late-time spectra ($\tau\gtrsim 488$\,d) of SN 2022erq are dominated by host-galaxy light. They show emission lines overlay H$\beta$ and H$\gamma$ absorption, indicating contributions from ionized gas and a substantial stellar population.  The spectrum at $\tau\approx 488$\,d was selected for its high signal-to-noise ratio (S/N) and high resolution to probe the local environment. Applying the R23 index \citep{2004ApJ...617..240K}, defined as ([\OII] $\lambda$3727 + [\OIII] $\lambda\lambda$4959, 5007)/H$\beta$, to this spectrum gives 12 + log(O/H) $= 8.45 \pm 0.05$, or $\sim 58$\% of the solar value \citep{Asplund2009}. This indicates a higher gas-phase metallicity for SN 2022erq than that of SN 2007if, derived with the same method ($\sim20$\% solar; \citealp{2011ApJ...733....3C}). 

To characterize integrated stellar properties, we fit the host SED using Pan-STARRS $g,r,i,z$ photometry and BayeSED3 \citep{2023ApJS..269...39H}. The modeling adopts \citet{BruzualG2003a} single stellar population models, a Chabrier initial mass function, an exponentially declining star-formation history, the \citet{CalzettiD2000a} dust-attenuation law, and nebular emission for populations younger than 10 Myr \citep{BylerN2017a}. 

\begin{figure}
 \centering
 \includegraphics[width=8.3cm,angle=0]{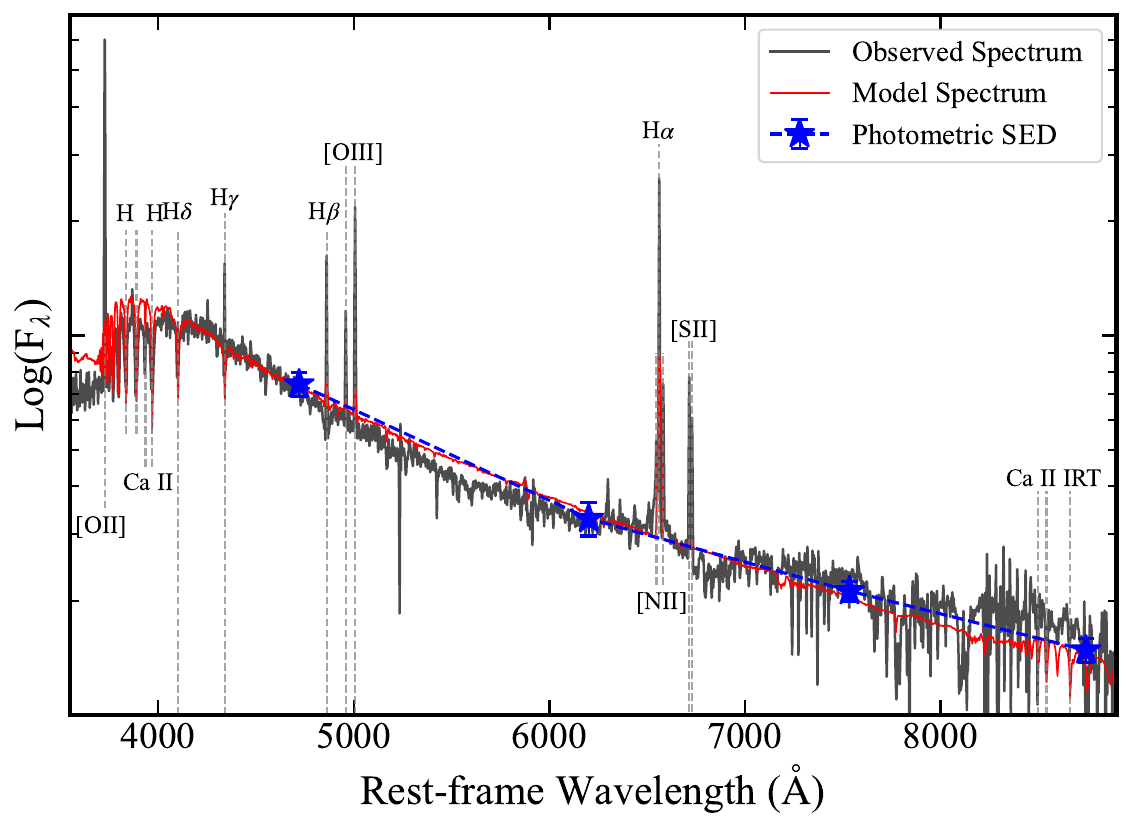}
 \caption{A comparison of the preexplosion host-galaxy photometric SED, the SED-derived model spectrum of the stellar population, and the late-time spectrum of SN 2022erq at $\tau\approx 488$ d (continuum-corrected to match the host SED). Dashed lines mark rest-frame wavelengths of spectral lines.} 
 \label{fig:host}
\end{figure}

\begin{figure*}
 \centering
 \includegraphics[width=17.6cm,angle=0]{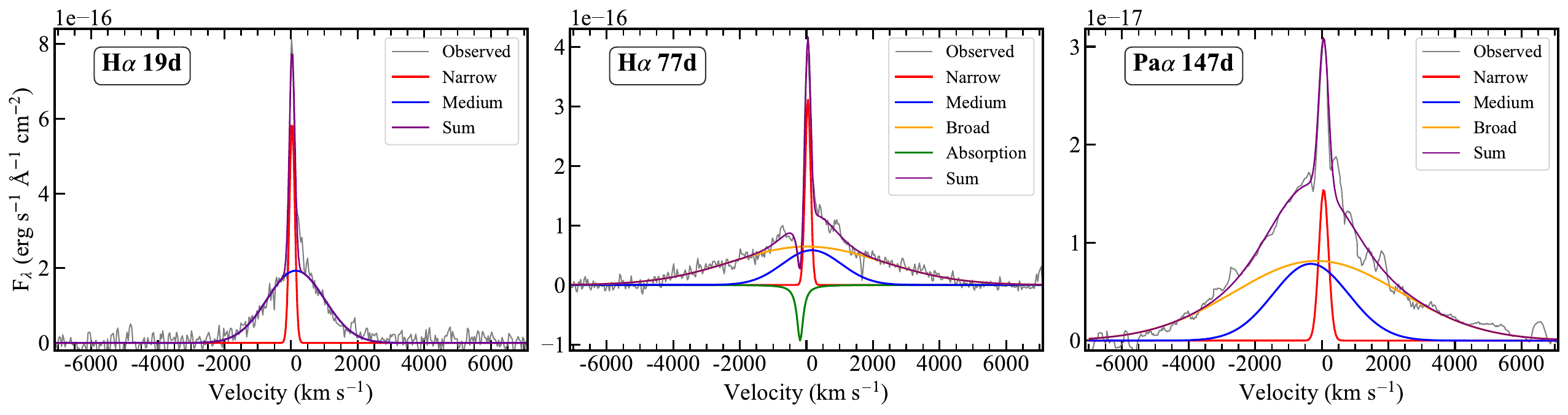}
 \caption{Multi-Gaussian fits to the H$\alpha$ and Pa$\alpha$ lines of SN 2022erq. The spectra with higher spectral resolution and S/N were selected to enable a robust decomposition of the line profiles. The instrumental FWHM is $\sim 210$\,km~s$^{-1}$ for the optical spectra and $\sim 150$\,km~s$^{-1}$ for the NIR spectrum.} 
 \label{fig:MCF}
\end{figure*}

The SED fit  constrains the stellar population age to $\log(t_{\rm age}/{\rm yr}) = 8.02^{+0.04}_{-0.02}$, corresponding to $\sim 105$\,Myr, indicating a relatively young stellar population.  The stellar metallicity is constrained to $\log(Z/{\rm Z}_{\odot}) = -2.13^{+0.16}_{-0.09}$, about 0.7\% of the solar metallicity, typical of low-metallicity galaxies. The dust attenuation in the $V$ band is $A_V = 0.046^{+0.08}_{-0.03}$ mag, indicating minimal dust obscuration. The stellar mass is $\log(M_{\star}/{\rm M}_{\odot}) = 7.97^{+0.03}_{-0.04}$, corresponding to about $9.3 \times\,10^7\,{\rm M}_{\odot}$, with an average star-formation rate over the past 100\,Myr of $\log({\rm SFR}_{100\,{\rm Myr}}/[{\rm M}_{\odot}\,{\rm yr}^{-1}]) = -0.17^{+0.16}_{-0.37}$, or  $\sim 0.67\,{\rm M}_{\odot}\,{\rm yr}^{-1}$.

A comparison between the pure stellar spectrum from the best-fit SED model and the observed host spectrum (Figure \ref{fig:host}) shows general agreement in Balmer and \CaII\ absorption. The prominent discrepancy lies in the strong gas emission lines (e.g., from O, S, N) present in the observation but not in the stellar population model. This highlights the different astrophysical origins of the two metallicity estimates: the R23 index probes the present-day, gas-phase metallicity, while the SED-derived value represents the average metallicity of the stellar population. Such a disparity, where recently enriched gas is more metal-rich than the older stellar bulk, is plausible in dwarf galaxies with recent star formation \citep{2006ApJ...636..214V, 2013ARA&A..51..457N,2023ApJ...948...52S}. 

{These results show that SN 2022erq resides in a low-luminosity, star-forming, dwarf galaxy with subsolar metallicity. This environmental profile is consistent with the established properties of SNe~Ia-CSM host galaxies \citep{2007arXiv0706.4088P, 2013ApJS..207....3S}. }

\begin{figure}
 \centering
 \includegraphics[width=8.3cm,angle=0]{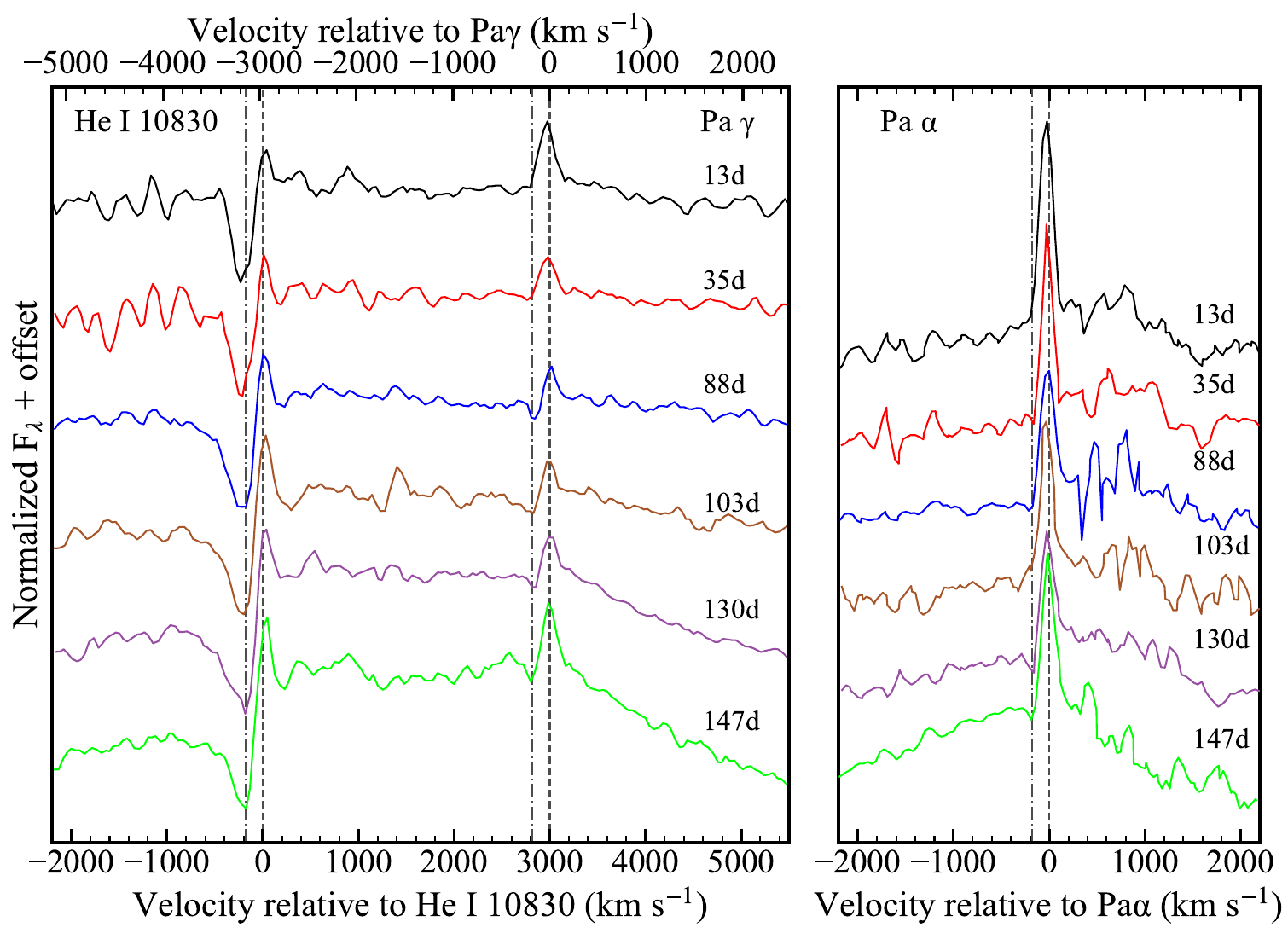}
\caption{Spectral features normalized and displayed in the velocity space at selected phases. The left panel shows the He\,I $\lambda10,830$ and Pa$\gamma$ lines, while the right panel presents the Pa$\alpha$ line. Vertical lines indicate velocities of $0$\,km~s$^{-1}$ (dashed) and $-180$\,km~s$^{-1}$ (dash-dotted) relative to host galaxy.} 
 \label{fig:HaHe}
\end{figure}

\subsection{{CSM Properties from the H$\alpha$ Emission}}
\label{csm}

The evolution of the H$\alpha$ profile provides a direct probe of the CSM properties and the interaction dynamics. 
Initially, the limited spectral resolution blends the instrumentally broadened narrow H$\alpha$ line with a weak, underlying broader component. As a result, the profile appears as a single feature and the broad base is difficult to discern.
At $\tau\approx 19$\,d (Figure \ref{fig:MCF}), improved resolution clearly decomposes the line into a narrow component ($\sim 200$\,km~s$^{-1}$) from ionized, unshocked CSM, and an intermediate component ($\sim 2000$\,km~s$^{-1}$), likely due to electron scattering \citep{2001MNRAS.326.1448C}. The FWHM of the narrow component is still close to the instrumental broadening, indicating a smaller intrinsic width (i.e., $\sim 100$\,km~s$^{-1}$ after the instrumental FWHM correction).

A distinct P-Cygni profile emerges by $\tau\approx 77$\,d (Figure~\ref{fig:MCF}). The velocity shift of its absorption minimum indicates a CSM wind velocity of $\sim 180$\,km~s$^{-1}$. The broad component further separates into an intermediate-width feature (FWHM $\approx 2000$\,km~s$^{-1}$) and a very broad component (FWHM $\approx 5400$\,km~s$^{-1}$) arising from the CDS formed between the forward and reverse shocks \citep{2001MNRAS.326.1448C}. The FWHM of this CDS component in SN 2022erq is significantly smaller than in SNe IIn ($15{,}000$--$20{,}000$\,km~s$^{-1}$; \citealt{2012ApJ...756..173S}) but consistent with other SNe~Ia-CSM (e.g., SNe 2005gj, 2018evt), suggesting systematically different CSM density distributions.

The NIR spectral evolution corroborates the findings from H$\alpha$. Both the He\,I $\lambda$10,830 line and the Paschen lines develop P-Cygni profiles with velocity separations similar to H$\alpha$ (Figure \ref{fig:HaHe}), and all eventually show broad underlying components with comparable widths (Figure~\ref{fig:MCF}), indicating a common emitting region. A key difference is the earlier emergence ($\tau\approx 13$\,d) of the narrow absorption component in the He\,I line compared to the hydrogen lines. This suggests that helium is more susceptible to resonance scattering, leading to detectable absorption features at earlier phases.

The mass-loss rate can be estimated from the narrow H$\alpha$ luminosity ($L_{\rm H\alpha}$), using $L_{\mathrm{H}\alpha} \approx 2\times10^{39}\,\dot{M}_{0.01}^{2}\,v_{\mathrm{w},500}^{-2}\,\beta\,(r/10^{15}\,\mathrm{cm})^{-1}\ \mathrm{erg\,s^{-1}}$ \citep{2013ApJ...768...47O}, which assumes a wind-density profile and recombination-dominated H$\alpha$ emission. Since the narrow H$\alpha$ emission arises from unshocked, photoionized CSM ahead of the forward shock, it mainly probes the outer CSM and thus mass loss at earlier preexplosion epochs.
  
Adopting a wind velocity of $180$\,km~s$^{-1}$ and the host-subtracted $L_{\rm H\alpha}$ measured at $\tau\approx 19$ and $77$\,d, we derive mass-loss rates of $4.5\times10^{-2}$ and $3.7\times10^{-2}$\,$M_\odot$\,yr$^{-1}$ for SN 2022erq, respectively, with an average value of $(4.1\pm0.4)\times10^{-2}$\,$M_\odot$\,yr$^{-1}$. This rate is intermediate compared to other SN Ia-CSM progenitors using an identical methodology \citep{2013ApJ...770...29C, 2023ApJ...948...52S}.

With interaction persisting for $\sim 400$\,d and an ejecta velocity of $10{,}000$\,km\,s$^{-1}$ derived from IGE absorption velocities, the CSM extends to $\sim 3.5\times10^{16}$\,cm. A steady wind of $180$\,km\,s$^{-1}$ would require $\sim 60$\,yr to build this shell. Combining this timescale with the average mass-loss rate derived from H$\alpha$ ($\dot{M} \approx 0.04\,M_\odot\,\mathrm{yr}^{-1}$) yields a total CSM mass of $\sim 2.4\,M_\odot$ for a spherical geometry (or less if the CSM is aspherical or clumpy).

\subsection{{CSM Properties from  Bolometry}}
\label{csm2}
The narrow H$\alpha$ emission originates from the outer, cool, and tenuous CSM, and its analysis yields the mass-loss rate at earlier epochs (several decades before the explosion).  Meanwhile, the post-peak evolution of the bolometric luminosity reflects the CSM density in the interaction region, i.e., the density at the location of the CDS formed by the ejecta-CSM interaction \citep{1994ApJ...420..268C,2004ApJ...616..339W,2013MNRAS.435.1520M}. Therefore, the CSM density derived from the bolometric light curve, together with that obtained from the H$\alpha$ emission, can jointly depict a more complete picture of the CSM density structure.

\begin{figure}
 \centering
 \includegraphics[width=8.3cm,angle=0]{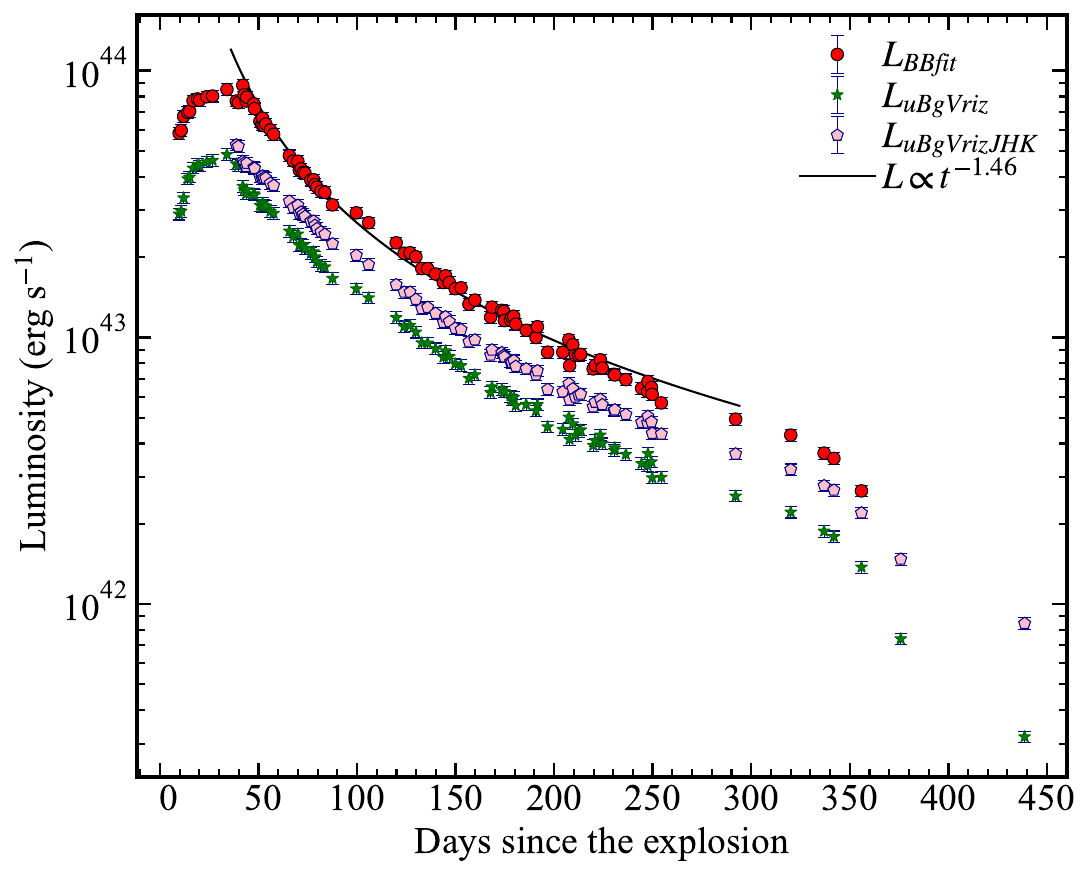}
 \caption{Bolometric light curve of SN~2022erq. Bolometric luminosities derived from blackbody fits to the observed SED are shown together with two sets of quasi-bolometric luminosities obtained by direct flux integration over the $uBgVriz$ and $uBgVrizJHK$ bands, respectively. The solid line represents the time-dependent power-law fit $L \propto t^{\alpha}$ to the post-peak light curve.}
 \label{fig:bolo_fit}
\end{figure}

Figure~\ref{fig:bolo_fit} shows the bolometric light curve of SN~2022erq, derived from blackbody fits to the observed SED spanning the $u$–$K$ bands. Light curves in individual filters were interpolated onto a common time grid, and blackbody fits were performed only when at least four filters had detections; no extrapolation was applied to epochs with missing bands. The figure also shows quasi-bolometric luminosities from direct flux integration over $uBgVriz$ and $uBgVrizJHK$. Although $u$-band data are unavailable for $t \gtrsim 40$ d, its contribution to the total luminosity is negligible at these epochs.

\begin{figure}
 \centering
 \includegraphics[width=8.3cm,angle=0]{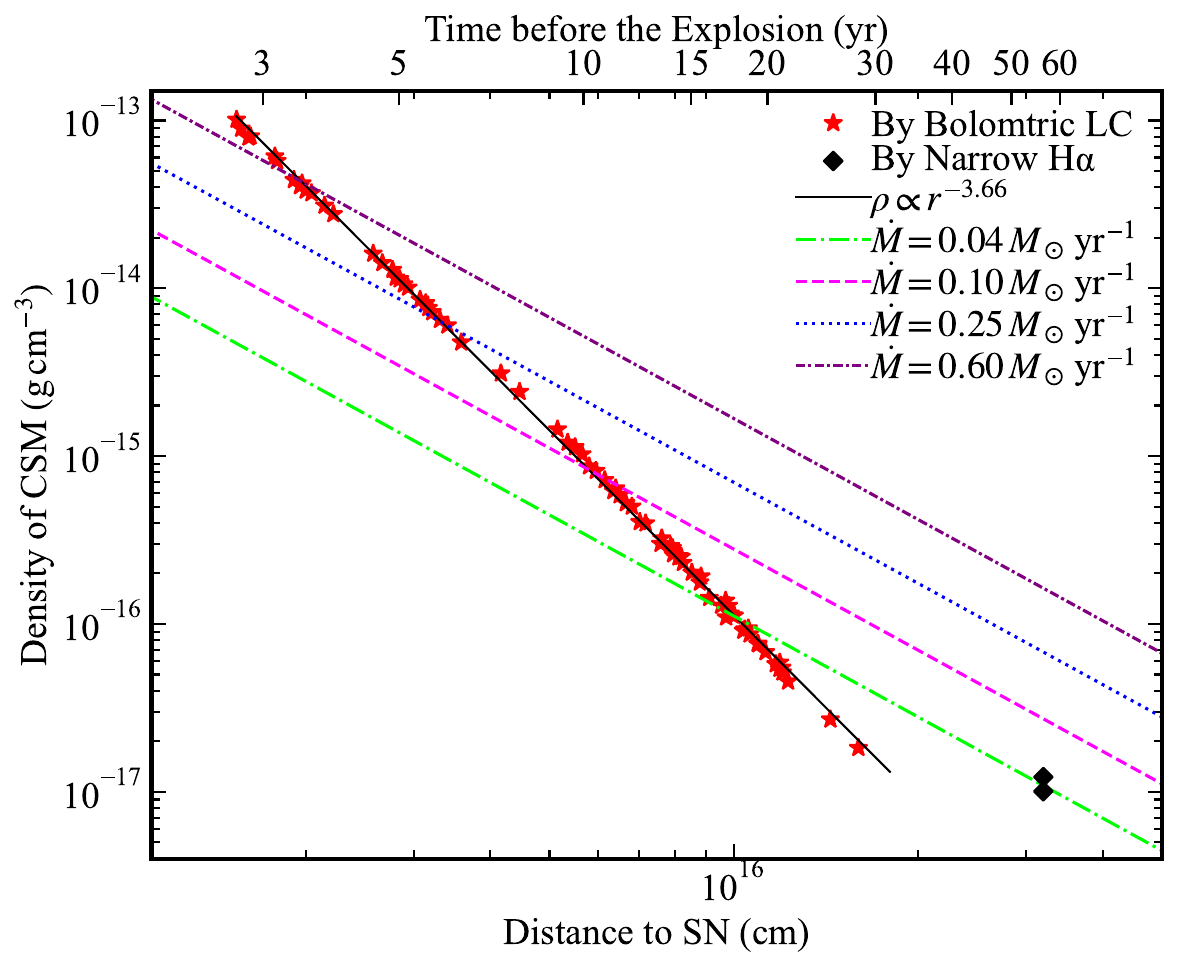}
\caption{CSM density profile of SN~2022erq. Red stars show densities inferred from the bolometric light curve, with a power-law fit $\rho_{\rm CSM}\,\propto r^{-s}$ (black line). Black diamonds mark independent density estimates from the narrow H$\alpha$ line flux. Coloured dash/dotted curves show steady-wind profiles $\rho=\dot{M}/(4\pi r^{2}v_{\rm wind})$ for different mass-loss rates, assuming a constant wind velocity $v_{\rm wind}=180\ {\rm km\ s^{-1}}$. The top axis gives the corresponding time before the explosion, $t = r / v_{\rm wind}$.}
 \label{fig:density}
\end{figure}

SN~2022erq reaches a peak bolometric luminosity of $L_{\rm peak} \approx 8 \times 10^{43}\ {\rm erg\ s^{-1}}$ at $\sim 30$ d after explosion, nearly an order of magnitude above typical SNe Ia. Applying Arnett's law \citep{1982ApJ...253..785A,2005A&A...431..423S} to this luminosity and rise time would require an implausible $^{56}$Ni mass of $M_{\rm Ni} \approx 6\,M_{\odot}$ for a thermonuclear SN, ruling out $^{56}$Ni decay as the dominant power source. The slow post-peak decline and clear spectroscopic signs of interaction instead indicate that shock interaction with dense CSM is powering the light curve.

In this interaction-powered picture, the kinetic energy dissipated at the forward shock is
\begin{equation}
  \mathrm{d}E_{\mathrm{kin}} 
  = 4\pi r_{\mathrm{sh}}^{2} \,\frac{1}{2}\rho_{\mathrm{CSM}} v_{\mathrm{sh}}^{2} \,\mathrm{d}r_{\mathrm{sh}},
\end{equation}
so that the bolometric luminosity can be written as
\begin{equation}
\label{eq2}
  L = \varepsilon \frac{\mathrm{d}E_{\mathrm{kin}}}{\mathrm{d}t}
    = 2\pi \varepsilon \,\rho_{\mathrm{CSM}}\, r_{\mathrm{sh}}^{2} v_{\mathrm{sh}}^{3},
\end{equation}
where $\varepsilon$ is the kinetic-to-radiative conversion efficiency, and $v_{\mathrm{sh}}$ is the velocity of the CDS. 

Assuming canonical ejecta parameters for a Chandrasekhar-mass SN~Ia ($M_{\rm ej} \sim 1.4\,M_\odot$, $v \sim 10^{4}$~km~s$^{-1}$), the total shock kinetic energy is $\sim 1.4$~foe. The total radiated energy obtained by integrating the bolometric light curve is $\sim 0.7$~foe, which implies $\varepsilon \approx 50\%$. 
The CDS velocity $v_{\mathrm{sh}}(t)$ is measured from the FWHM of the broad H$\alpha$ component: $v_{\mathrm{sh}} \approx 5400$~km~s$^{-1}$ at $\tau \approx 77$~d and $5900$~km~s$^{-1}$ at $\tau \approx 147$~d (Figure~\ref{fig:MCF}), which can be parameterized as $v_{\mathrm{sh}}(t) = v_1 \, t_d^{0.15}$, where $t_d$ is the time in days since explosion and $v_1 = 2800$~km~s$^{-1}$. The shock radius $r_{\mathrm{sh}}(t)$ is then obtained by integrating $v_{\mathrm{sh}}(t)$ over time.

With $L(t)$, $v_{\mathrm{sh}}(t)$, and $r_{\mathrm{sh}}(t)$, we can invert the above relation to derive the CSM density $\rho_{\mathrm{CSM}}(r)$. The interaction-dominated phase begins at $\tau \approx 40$~d, when the luminosity has passed its peak and started to decline while remaining about an order of magnitude above a typical SN~Ia, and when a pronounced CDS component appears in the spectra. The interaction-dominated phase is taken to end when the shock reaches the inner ejecta, corresponding to the rapid decline of the light curve at $\tau \approx 330$~d. The inferred CSM densities are shown as red stars in Figure~\ref{fig:density}. The CSM traced by the density profile extends from $\sim 1.5\times10^{15}$ cm to $\sim 1.6\times10^{16}$ cm, corresponding to a total CSM mass of $\sim 3.0$~$M_\odot$. For comparison, independent density estimates derived from the narrow H$\alpha$ emission are also plotted in Figure~\ref{fig:density}, constraining the outer CSM properties. The two sets of measurements are broadly consistent, both indicating a decreasing density at larger radii (i.e., at earlier epochs).

A power-law fit to these numerical results yields $\rho_{\mathrm{CSM}} \propto r^{-3.66}$. As an independent check, we fit the post-peak bolometric luminosity with $L \propto t^{\alpha}$, obtaining $\alpha = -1.46$. Using the analytic model of \citet{2013MNRAS.435.1520M}, the CSM density is parameterized as $\rho_{\mathrm{CSM}} \propto r^{-s}$, and the light-curve slope is related to the CSM density exponent $s$ by $\alpha = \frac{6s-15+2n-ns}{n-s}$, where $n \simeq 10$ is expected for SN~Ia ejecta \citep{Matzner_1999,2010ApJ...708.1025}. From this, we derive $s = 3.59$, corresponding to $\rho_{\mathrm{CSM}} \propto r^{-3.59}$, in good agreement with the numerical result. 

For a steady wind with constant mass-loss rate and wind velocity $v_{\rm wind}$, the density profile is $\rho = \dot{M} / (4\pi r^{2} v_{\rm wind})$, i.e., $\rho_{\mathrm{CSM}} \propto r^{-2}$. Taking $v_{\rm wind} = 180$~km~s$^{-1}$ (as derived in Sect.~\ref{csm}), this steady-wind behaviour is illustrated by the coloured dashed lines in Figure~\ref{fig:density}. Compared with a steady wind, the much steeper density profile of SN~2022erq implies a faster decline of density with radius, suggesting that the progenitor lost increasingly more material as it approached explosion. Starting about 20~yr before explosion, the mass-loss rate reached $\sim 0.04$~M$_\odot$~yr$^{-1}$, increased to $\sim 0.1$~M$_\odot$~yr$^{-1}$ around 10~yr before explosion, and rose further to $\sim 0.25$~M$_\odot$~yr$^{-1}$ about 5~yr prior. In the final $\sim 3$~yr, it escalated to an extraordinary $\sim 0.6$~M$_\odot$~yr$^{-1}$. This trend indicates a dramatic enhancement of mass loss as the system approached explosion. These timescales are derived assuming a constant wind speed; if the outflow velocity evolved with time, the absolute timing would change, and the values quoted here should be regarded as approximate.

\subsection{{Progenitor System}}
\label{proge}

The properties of SN~2022erq indicate a thermonuclear explosion inside a massive hydrogen-rich CSM. The inferred preexplosion mass-loss rate ($\sim0.04$–$0.6$~M$_\odot$~yr$^{-1}$ over decades) exceeds by orders of magnitude the steady rates predicted by conventional SD models ($\sim 10^{-5}$–$10^{-4}$~M$_\odot$~yr$^{-1}$; e.g., \citealp{1999ApJ...522..487H,2009PASJ...61.1251M}). Nevertheless, the presence of such a massive hydrogen-rich CSM points to a progenitor with a nondegenerate companion, most likely an AGB star or a moderate-mass main-sequence star, consistent with enhanced-accretion scenarios for dense CSM formation \citep{2006MNRAS.368.1095H}. However, even the common-envelope wind (CEW) model \citep{2018ApJ...861..127M} and the CD scenario \citep{2011arXiv1109.4652S,2017MNRAS.464.3965W}, which allow larger CSM reservoirs ($\sim 0.1$–2~M$_\odot$), still marginally fall short of the $\sim 3$~M$_\odot$ of CSM inferred for SN~2022erq and it remains challenging for these models to naturally produce mass-loss rates as extreme as $\sim 0.6$~M$_\odot$~yr$^{-1}$. Instead, this high rate likely requires a brief, violent ejection episode shortly before explosion, for example a delayed dynamical instability in the CEW framework (see Figure~2 in \citealt{2018ApJ...861..127M}) or a violent prompt merger in the CD scenario.

The young host environment and extensive CSM of SN~2022erq further support a short progenitor evolutionary timescale with substantial mass loss, consistent with a close binary that experienced a brief, high-mass-transfer phase immediately before explosion. The donor could be an intermediate-mass star ($\sim 3$–7~M$_\odot$) undergoing rapid Roche-lobe overflow or incipient common-envelope evolution, although a lower-mass companion (for example, an AGB or red giant  star) entering a dynamically unstable phase is also plausible. The short delay time implied by the young host also aligns with CD scenarios in which a WD rapidly merges with the core of a giant companion \citep{2017MNRAS.464.3965W}, although this channel faces additional constraints on retaining hydrogen-rich material.

\section{{Summary}}
\label{summ}
SN~2022erq is the earliest confirmed member of the rare Ia-CSM class, a superluminous thermonuclear transient displaying clear circumstellar interaction signatures within two days of explosion. Through a combined analysis of early-time spectroscopy and densely sampled photometry, this work shows that its extraordinary properties are driven by a massive, extended, hydrogen-rich CSM.

Photometrically, SN~2022erq reached a peak bolometric luminosity of
$L_{\rm peak} \approx 8 \times 10^{43}\ \mathrm{erg\ s^{-1}}$, exceeding that of
normal SNe~Ia by roughly an order of magnitude. This extreme luminosity,
together with its exceptionally slow post-peak decline, point to a
dominant energy source beyond radioactive decay. 

Spectroscopically, the early phases are dominated by IGEs with weak IME features, overlaid with persistent H$\alpha$ emission from CSM interaction. At later epochs, the emergence and long-lived presence of a broad H$\alpha$ component indicate sustained shock interaction and the formation of a CDS, which is also accompanied by a plateau in the colour curve. In particular, the progressive strengthening of the H$\alpha$ flux relative to the continuum contributes additional flux in the $r$ band and helps to keep the $g-r$ colour relatively red, rather than reflecting an intrinsic trend of the SED becoming redder.

These observations indicate that SN~2022erq arose from a highly efficient thermonuclear explosion embedded in an exceptionally massive, hydrogen-rich CSM. The IGE-rich, IME-weak spectrum points to efficient burning broadly similar to that inferred for some overluminous SNe~Ia, while the peak and sustained superluminous output far exceeds what can be powered by radioactive decay alone. Instead, the light-curve morphology, color evolution, and H$\alpha$ diagnostics together show that the observed radiation is overwhelmingly powered by ejecta–CSM shock interaction. In this picture, the extraordinary luminosity of SN~2022erq is a direct consequence of the efficient conversion of mechanical energy into radiation within the dense CSM, rather than a fundamentally different explosion mechanism.

Analysis of the H$\alpha$ profile reveals a preexplosion mass-loss history characterized by a wind velocity of $\sim 180$~km~s$^{-1}$ and a rate of $\sim 0.04$~M$_\odot$~yr$^{-1}$ sustained for $\sim 60$~yr, producing a $\sim 2.4$~M$_\odot$ CSM shell extending to $\sim 3.5\times10^{16}$~cm. The bolometric light curve further constrains the evolution of the mass loss in the final decades before explosion, indicating that the mass-loss rate increased by an order of magnitude, from $\sim 0.04$ to $\sim 0.6$~M$_\odot$~yr$^{-1}$ as the system approached explosion. Integrating the CSM density profile inferred from the bolometric light curve yields a total CSM mass of $\sim 3.0$~M$_\odot$. The two mass estimates are broadly consistent. The lower value from H$\alpha$ is expected, as it primarily traces the outer, ionized CSM under the assumption of a constant mass-loss rate, while the bolometric modeling probes the denser inner CSM and thus likely provides a more accurate measure of the total CSM mass.

SN~2022erq therefore demonstrates that understanding extreme thermonuclear transients requires disentangling the progenitor’s explosion mechanism from its preexplosion mass-loss history. The escalation of mass loss in the final decades places strong constraints on the progenitor system and its late-stage evolution. In summary, the combined constraints—a massive hydrogen-rich CSM formed on decade-to-century timescales, a luminous thermonuclear event, and a young host environment—favor a close-binary progenitor that underwent a short-lived, high-mass-loss episode immediately before explosion. The most natural interpretation involves a near‑Chandrasekhar‑mass WD and a nondegenerate companion of intermediate mass, consistent with the mass‑transfer and ejection timescales inferred from the CSM. This picture challenges steady-wind SD models and highlights the need for further theoretical and observational work on short-timescale, interaction-driven progenitor channels. Future high-cadence, high-resolution spectroscopy ($R\gtrsim 5000$), complemented by polarimetry and multi-wavelength campaigns, will be essential for constraining the CSM geometry, explosion asymmetry, and progenitor nature in such strongly interacting systems.

\begin{acknowledgments}
  
This work is supported by the National Key R\&D Program of China with grant 2021YFA1600404, the B-type Strategic Priority Program of the Chinese Academy of Sciences (grant XDB1160202), the National Natural Science Foundation of China (NSFC grants 12173082, 12333008,  and 12225304), the Yunnan Fundamental Research Projects (YFRP grants 202501AV070012, 202401BC070007, 202501AS070005  and  202605AS350010), the Top-notch Young Talents Program of Yunnan Province, the Light of West China Program provided by the Chinese Academy of Sciences, the Yunnan Revitalization Talent Support Program (Yunling Scholar Project), and the International Centre of Supernovae, Yunnan Key Laboratory (grant 202302AN360001). X.F. Wang is supported by the NSFC (grants 12288102, 12033003, and 11633002) and the Tencent Xplorer Prize. 
A.V.F.’s research group at U.C. Berkeley acknowledges financial assistance from the Christopher R. Redlich Fund, as well as donations from Gary and Cynthia Bengier, Clark and Sharon Winslow, Alan Eustace and Kathy Kwan, Timothy and Melissa Draper, Briggs and Kathleen Wood, Ellyn and Alan Seelenfreund (W.Z. is a Bengier-Winslow-Eustace Specialist in Astronomy, T.G.B. is a Draper-Wood-Seelenfreund Specialist in Astronomy), and numerous other donors. 
Y.-Z. Cai is supported by the NSFC (Grant No. 12303054), the Yunnan Fundamental Research Projects (Grant Nos. 202401AU070063, 202501AS070078), the National Key Research and Development Program of China (Grant No. 2024YFA1611603). Y.-Z. Cai also acknowledges financial support from the SOXS project (PI S. Campana).

A.P., A.R., N.E.R., and G.V. acknowledge support from the PRIN-INAF 2022, ``Shedding light on the nature of gap transients: from the observations to the models.''
A.R. also acknowledges financial support from the GRAWITA Large Program Grant (PI P. D'Avanzo).
N.E.R. also acknowledges support from the Spanish Ministerio de Ciencia e Innovaci\'on (MCIN) and the Agencia Estatal de Investigaci\'on (AEI) 10.13039/501100011033 under the program Unidad de Excelencia Mar\'ia de Maeztu CEX2020-001058-M.
J.V. is supported by the Hungarian NKFIH-OTKA grant K142534. The operation of the RC80 telescope at Konkoly is based on the GINOP 2.3.2-15-2016-00033 grant funded by the European Union.
C.P.G. acknowledges financial support from the Secretary of Universities and Research (Government of Catalonia) and by the Horizon 2020 Research and Innovation Programme of the European Union under the Marie Sk\l{}odowska-Curie and the Beatriu de Pin\'os 2021 BP~00168 programme. 
S. Moran is funded by Leverhulme Trust grant RPG-2023-240. 
G.V. and I.S. acknowledge financial support from the SOXS project (PI S. Campana).
M.D.S. is funded by the Independent Research Fund Denmark (IRFD, grant  10.46540/2032-00022B) and by an Aarhus University Research Foundation Nova project (AUFF-E-2023-9-28).
D.D.S. acknowledges support from the NSFC (grants 12303015 and 12273118), the National Science Foundation of Jiangsu Province (BK20231106), and the China Manned Space Program with grant  CMS-CSST- 2025-A20. T.M.R is part of the Cosmic Dawn Center (DAWN), which is funded by the Danish National Research Foundation under grant DNRF140. T.M.R acknowledges support from the Research Council of Finland project 350458.
L.G. acknowledges financial support from MCIN and AEI 10.13039/501100011033 under projects PID2023-151307NB-I00, CEX2020-001058-M, and by the MaX-CSIC Excellence Award MaX4-SOMMA-ICE.

We thank S. Taubenberger for his observation on 2022-07-01. We acknowledge the support of the staff of the LJT, XLT, NOT, TNG, GTC, Keck II, KOT, TNT, LT,  WMS and HET. Funding for the LJT has been provided by the CAS and the People's Government of Yunnan Province. The LJT is jointly operated and administrated by YNAO and the Center for Astronomical Mega-Science, CAS. We thank the observers who shared their Keck-II NIR observations on WISeREP.

Based in part on observations made with the Nordic Optical Telescope, owned in collaboration by the University of Turku and Aarhus University, and operated jointly by Aarhus University, the University of Turku, and the University of Oslo, representing Denmark, Finland, and Norway, the University of Iceland, and Stockholm University, at the Observatorio del Roque de los Muchachos, La Palma, Spain, of the Instituto de Astrofisica de Canarias. Observations from the NOT were obtained through the NUTS2 collaboration which is supported in part by the Instrument Centre for Danish Astrophysics (IDA), and the Finnish Centre for Astronomy with ESO (FINCA) via Academy of Finland grant  306531. The data presented here were obtained in part with ALFOSC, which is provided by the Instituto de Astrofisica de Andalucia (IAA) under a joint agreement with the University of Copenhagen and NOTSA. The Liverpool Telescope is operated on the island of La Palma by Liverpool John Moores University in the Spanish Observatorio del Roque de los Muchachos of the Instituto de Astrofisica de Canarias with financial support from the UK Science and Technology Facilities Council. The Italian Telescopio Nazionale Galileo (TNG) operated on the island of La Palma by the Fundaci\'on Galileo Galilei of the INAF (Istituto Nazionale di Astrofisica) at the Spanish Observatorio del Roque de los Muchachos of the Instituto de Astrofísica de Canarias. Based in part on observations made with the Gran Telescopio Canarias (GTC) under the programme GTCMULTIPLE2A-22A,  installed at the Spanish Observatorio del Roque de los Muchachos of the Instituto de Astrofísica de Canarias, on the island of La Palma.
This article is also based in part on observations made in the Observatorios de Canarias del IAC with the Telescopio Nazionale Galileo, under the program A44TAC\_22 (PI G. Valerin).

Based in part on observations obtained with the Hobby-Eberly Telescope (HET), which is a joint project of the University of Texas at Austin, the Pennsylvania State University, Ludwig-Maximillians-Universitaet Muenchen, and Georg-August Universitaet Goettingen. The HET is named in honor of its principal benefactors, William P. Hobby and Robert E. Eberly. The Low Resolution Spectrograph 2 (LRS2) was developed and funded by the University of Texas at Austin McDonald Observatory and Department of Astronomy, and by Pennsylvania State University. We thank the Leibniz-Institut fur Astrophysik Potsdam (AIP) and the Institut fur Astrophysik Goettingen (IAG) for their contributions to the construction of the integral field units. We acknowledge the Texas Advanced Computing Center (TACC) at The University of Texas at Austin for providing high performance computing, visualization, and storage resources that have contributed to the results reported within this paper.

Some of the data presented herein were obtained at the W. M. Keck
Observatory, which is operated as a scientific partnership among the
California Institute of Technology, the University of California, and
NASA; the observatory was made possible by the generous financial
support of the W. M. Keck Foundation.
A major upgrade of the Kast spectrograph on the Shane 3 m
telescope at Lick Observatory, led by Brad Holden, was made possible through
generous gifts from the Heising-Simons Foundation, William and Marina Kast,
and the University of California Observatories. Research at Lick Observatory is
partially supported by a generous gift from Google.

\end{acknowledgments}

\appendix

\counterwithin{figure}{section}
\renewcommand{\thefigure}{\Alph{section}\arabic{figure}}

\section{Supplemental materials}
\label{app}

\begin{figure*}
 \centering
 \includegraphics[width=14cm,angle=0]{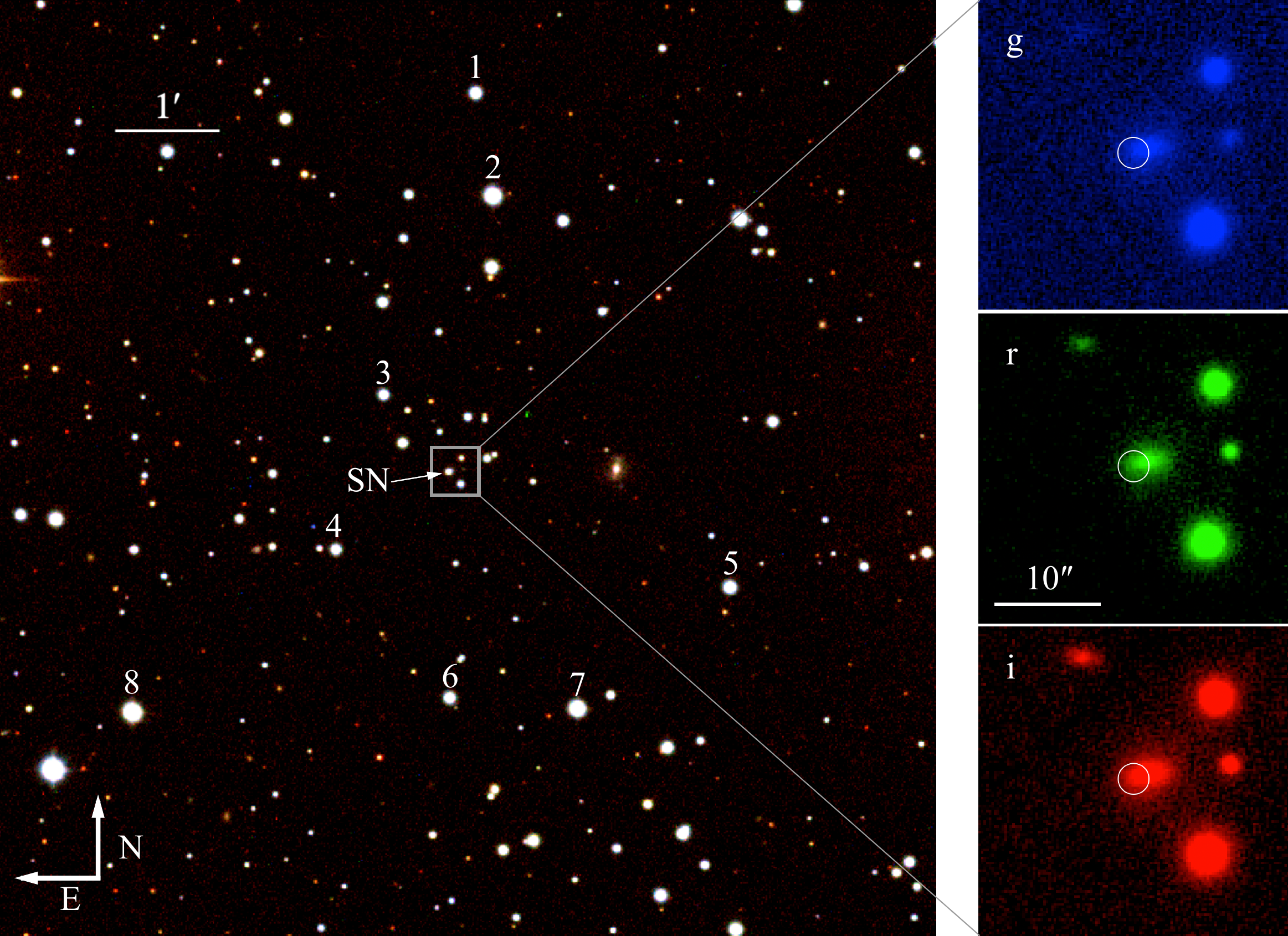}
 \caption{Composite $gri$- band finder chart of SN 2022erq and local reference stars obtained on June 2022 (left), alongside a preexplosion Pan-STARRS $gri$- band image (right). The position of SN is indicated by a white circle (radius = 1\farcs5), matching the typical image FWHM.
 }
 \label{<img>}
\end{figure*}

\begin{figure*}
 \centering
 \includegraphics[width=14cm,angle=0]{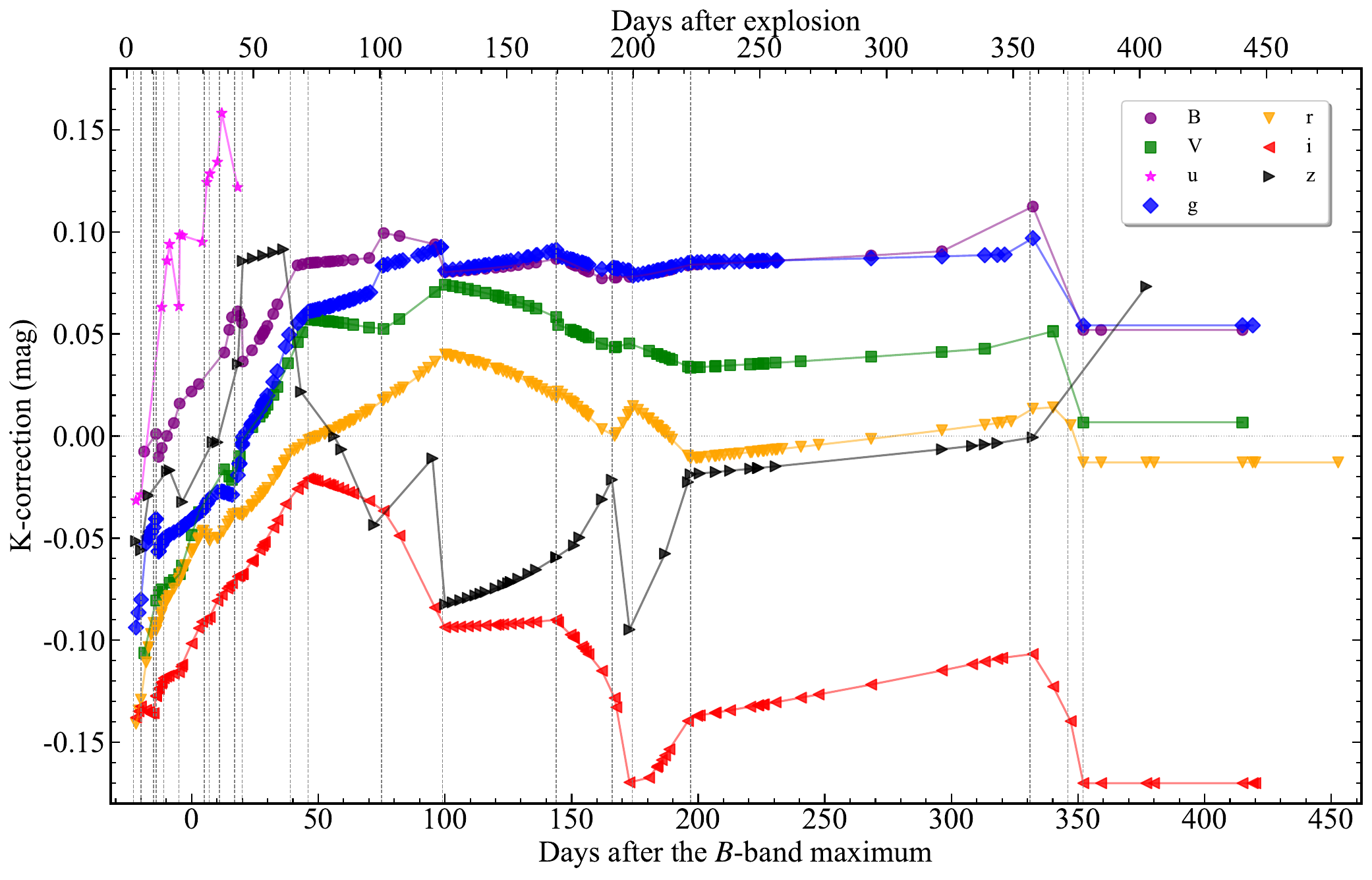}
 \caption{K-correction for the $BVugriz-$ band light curves of SN 2022erq, derived from spectra obtained at comparable phases. Epochs with spectroscopic observations are indicated by vertical dashed lines. }
 \label{<Kcor>}
\end{figure*}

\begin{figure*}
 \centering
 \includegraphics[width=15cm,angle=0]{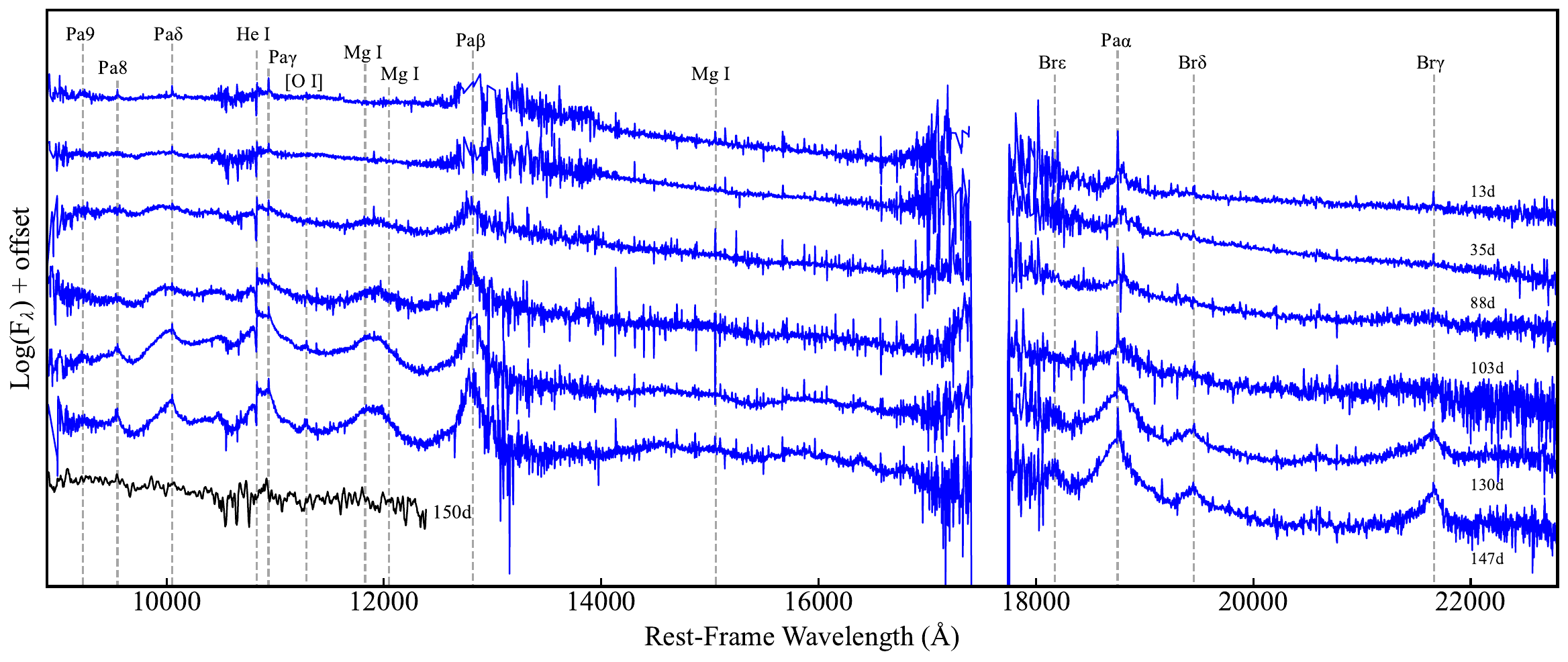}
 \caption{NIR spectra of SN 2022erq obtained by EMIR at GTC (black) and NIRES at Keck II (Blue). The public spectra of Keck are available through WISeREP. The dashed lines mark the narrow emission features at the rest wavelength. The phases of each spectrum are relative to the explosion date (MJD = 59649.1). The data behind this figure are available in machine-readable format in the online journal (see Figure~\ref{<sp>}).} 
 \label{fig:SpNIR}
\end{figure*}

This appendix compiles supplemental data and methods for SN 2022erq, including the finder chart, observing logs, instrument information, $K$-corrections, and the spectral comparison procedure. Table \ref{refStar} lists the local reference stars; Table~\ref{tab:photo} gives the $BVugriz$-band photometry; Table~\ref{app_a:JHK} presents the $JHK$-band photometry of SN 2022erq; and Table \ref{Tab:Spec_log} summarizes the spectroscopic observations.
Figure \ref{<img>} shows a finder chart for SN 2022erq and the local reference stars used in this study. Figure \ref{<Kcor>} presents the $K$-corrections. Figure~\ref{fig:SpNIR} displays the GTC and Keck II NIR spectra; the Keck II NIR data are publicly available via WISeREP (www.wiserep.org/object/20439; \citealp{2012PASP..124..668Y}). Figure~\ref{fig:Spmatch} illustrates the SN-component matching procedure.

\subsection{Facilities}
\label{app:Faci}

{Photometric observations were obtained by the  Li-Jiang 2.4\,m telescope (LJT; \citealp{2015RAA....15..918F}) and the Wumingshan 60-cm Telescope (WMS) of Yunnan Observatories, the Konkoly Observatory 0.8-m RC80 robotic  telescope (KOT), the Tsinghua-NAOC 0.8-m Telescope (TNT) at Xinglong Observatory, the Liverpool 2.0-m Telescope (LT), and the 67/91-cm Schmidt Telescope (ST). They were observed in the $JHK$- band by the 2.56-m Nordic Optical Telescope (NOT) at Roque de los Muchachos Observatory also. } 

Spectroscopic observations were carried out using various facilities, as follows.
\begin{itemize}
\item{Yunnan Faint Object Spectrograph and Camera (YFOSC; \citealp{2019RAA....19..149W}) on the LJT.}
\item{Beijing Faint Object Spectrograph and Camera (BFOSC) on the Xinglong 2.16\,m telescope (XLT) at the National Astronomical Observatories \citep{2016PASP..128j5004Z}. {Due to operational constraints, the slit could not be oriented at the parallactic angle.}}
\item{Kast Double Spectrograph (KDS) on the 3\,m Shane telescope at Lick Observatory \citep{miller1994lick}.}
\item{Alhambra Faint Object Spectrograph and Camera (ALFOSC) on the NOT.}
\item{Device Optimized for Low Resolution (DOLORES) with various grisms (low-resolution LRB/LRR and higher-resolution VHR) on the 3.58\,m Telescopio Nazionale Galileo (TNG) at Roque de los Muchachos Observatory.}
\item{Espectrógrafo Multiobjeto Infrarrojo (EMIR) on the 10.4\,m Gran Telescopio Canarias (GTC) at  Roque de los Muchachos Observatory.}
\item{Deep Extragalactic Imaging Multi-Object Spectrograph (DEIMOS) and Near Infrared Echellette Spectrograph (NIRES) on the 10\,m Keck II telescope at W. M. Keck Observatory.}
\item{Low Resolution Spectrograph-2 (LRS2; \citealp{2010SPIE.7735E..7HL}) on the 9.2\,m Hobby–Eberly Telescope (HET) at McDonald Observatory.}
\end{itemize}

LRS2 is an integral field spectrograph equipped with lenslets and a fiber array (commonly referred to as an integral field unit, or IFU). Given that the host galaxy of SN 2022erq has an optical extent of $\sim5\arcsec \times 6\arcsec$ and the SN is located at a projected angular distance of $\sim1.5\arcsec$ from the galactic center, the IFU's field of view of $7\arcsec \times 12\arcsec$ (with a fill factor close to 100\%) readily encompasses the entire galaxy and the SN position. The spectral features extracted specifically from the location of SN 2022erq are consistent with the spectrum from other regions of the galaxy. This indicates that, at the epoch of observation, the contribution from the SN itself is minimal, and the obtained spectrum is dominated by the integrated light of the host galaxy. Therefore, as an IFU-based spectrograph, LRS2 is an ideal facility for studying the environment of the SN.

The DEIMOS detector has a CCD array in which all spectra cross a small chip gap.  In the case of SN 2022erq, this chip gap is located right at the position of H$\alpha$ emission (from $\sim 6548$ to 6560\,\AA\ in the rest frame). This resulted in the loss of a segment of the narrow H$\alpha$ flux, as part of the line profile fell within the gap. For this reason, the narrow H$\alpha$ component from this spectrum ($\tau\approx 377$\,d) was not used for flux measurements and quantitative analysis.

\subsection{{Methodology for Spectral Comparison}}
\label{app:Spmatch}
We employed a custom spectral-matching algorithm to compare the spectra of SN~2022erq with those of well-observed SN~Ia templates \citep{2003Natur.424..651H,2006ApJ...650..510A}. The procedure automatically processes pairs of spectra over a wide range of post-explosion phases and identifies spectrally similar matches. To allow for the slower spectral evolution expected in strongly interacting events, we do not restrict ourselves to strictly contemporaneous phases, but instead permit phase differences of up to 120~days.

Each spectrum is first corrected for redshift and scaled to a common flux normalization by multiplying by $4\pi D^{2}$, where $D$ is the distance. The core of the matching process involves computing an initial difference between the SN~2022erq spectrum and a given template, and fitting this difference with a third-order polynomial as a smooth, wavelength-dependent correction to the continuum. Adding this polynomial to the template spectrum yields a “matched” spectrum for direct comparison with SN~2022erq.

\begin{figure*}
 \centering
 \includegraphics[width=16cm,angle=0]{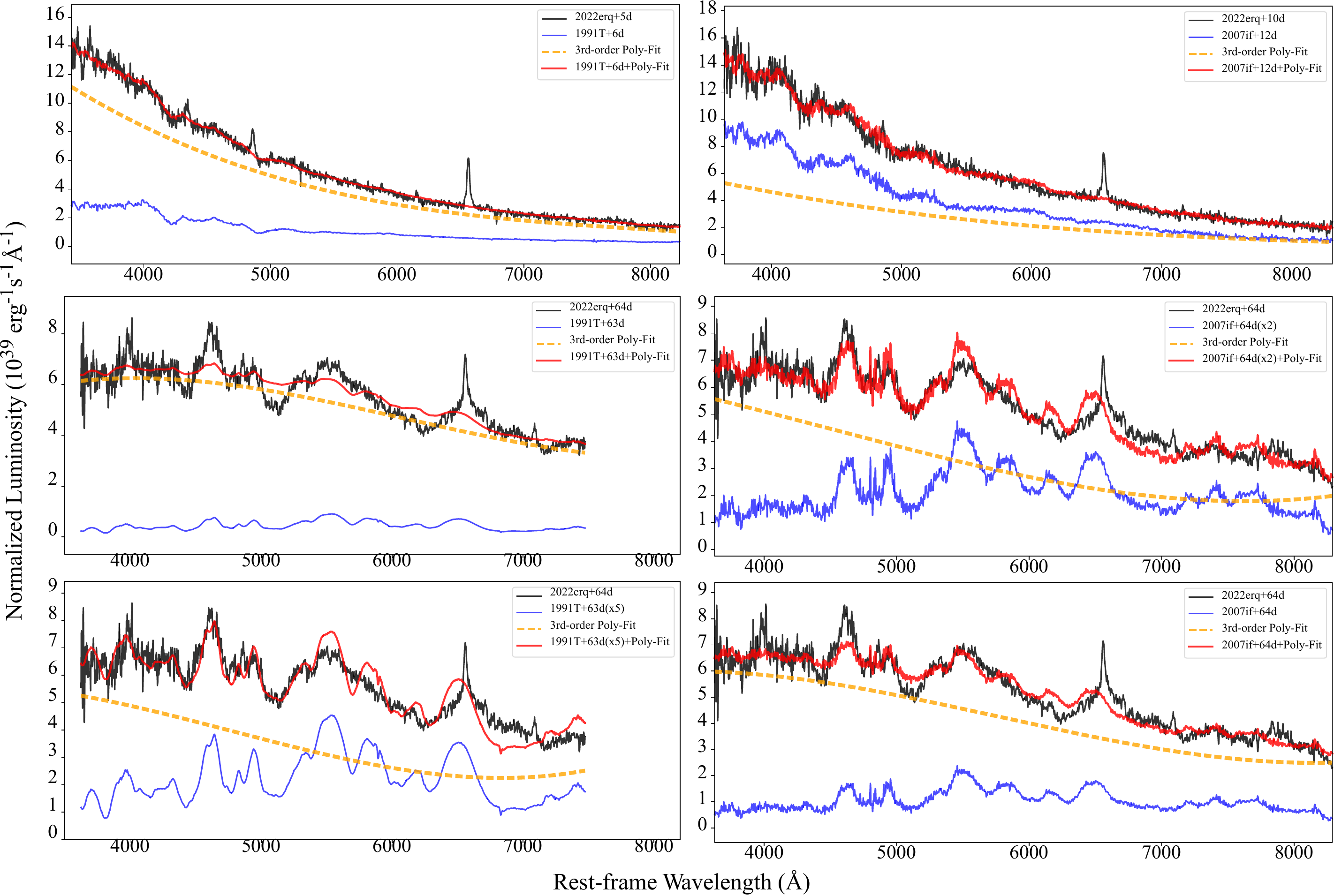}
 \caption{An illustration of the spectral matching procedure. The spectra of the comparison SNe were corrected for redshift and distance, and then combined with a third-order polynomial fit to match the corresponding continuum of SN 2022erq. For phases where the template luminosity differs significantly from that of SN 2022erq, such as $\tau>60$\,d in the case of SNe 1991T ($\times5$) and 2007if ($\times2$), an additional flux scaling is applied prior to polynomial fitting to preserve the strength of spectral features. } 
 \label{fig:Spmatch}
\end{figure*}

In cases where the luminosities of the template and SN~2022erq differ substantially, a polynomial continuum correction alone can dilute the spectral features of the matched template. To mitigate this, we first apply a global flux scaling to the template prior to the polynomial fitting. The scaling factor is chosen such that the flux level around relatively isolated Fe~{\sc ii} $\lambda\lambda4549,4629$ features approximately matches that of SN~2022erq. The third-order polynomial is then fitted to the residual difference after this scaling, and the sum of the scaled template and polynomial correction is taken as the final matched spectrum. As illustrated in Figure~\ref{fig:Spmatch}, this step is particularly important at $\tau \gtrsim 60$~d, where the luminosities of SN~1991T and SN~2007if are much lower than that of SN~2022erq.

The quality of each match is quantified by a set of residual-based metrics. Let $F_{\rm SN}(\lambda)$ be the flux of SN~2022erq and $F_{\rm temp,fit}(\lambda)$ the matched template (i.e., template plus polynomial correction). We define the absolute residual
\[
R(\lambda) = F_{\rm SN}(\lambda) - F_{\rm temp,fit}(\lambda),                                                                                                                              
\]
and the relative residual
\[
R_{\rm rel}(\lambda) = \frac{R(\lambda)}{|F_{\rm SN}(\lambda)| + \epsilon},
\]
where the small constant $\epsilon$ prevents division by zero. Over wavelength regions that exclude strong CSM emission lines and areas of severe blending, we compute for both $R$ and $R_{\rm rel}$ the mean, standard deviation, and rms. We also record the minimum, maximum, and range of $R$, as well as higher-order statistics such as the skewness and kurtosis of the residual distribution. Additional RMS quantities are computed for (i) the difference between $F_{\rm SN}$ and the uncorrected template plus polynomial fit, and (ii) the difference between the original template and the SN spectrum with the polynomial term subtracted, to assess the internal consistency of the continuum correction.

Among all candidate template spectra for a given SN~2022erq epoch, we use the RMS of the relative residual, $\mathrm{RMS}(R_{\rm rel})$, as our primary ranking statistic, and identify the three templates with the smallest values as the best matches. For these top-ranked matches, we further examine the absolute RMS residuals and phase differences and visually inspect the spectra to ensure that the selected matches are morphologically plausible.

We emphasize that, because the spectra of SN~2022erq are strongly affected by CSM interaction (continuum boosting and line dilution), even small differences in RMS residuals do not uniquely determine the intrinsic SN~Ia subtype. The spectral comparisons are therefore used in a heuristic sense: to demonstrate that SN~2022erq shares an efficient-burning spectroscopic phenotype (iron-group dominance and weak IMEs) with both 91T-like and some SC SNe~Ia, rather than to identify a single, uniquely ``best'' template.

\setcounter{table}{0} 
\renewcommand{\thetable}{\Alph{section}\arabic{table}}

\begin{deluxetable*}{lcccccccccc}[!th]
\label{refStar}
\tablecaption{Local reference star on the field of SN 2022erq.}
\tabletypesize{\scriptsize}
\tablewidth{0pt}
\tablehead{
\colhead{Ref.} &   \colhead{\textit{B}} & \colhead{\textit{V}}& \colhead{\textit{u}} &\colhead{\textit{g}}& \colhead{\textit{r}}& \colhead{\textit{i}}& \colhead{\textit{z}} & \colhead{\textit{J}}& \colhead{\textit{H}}& \colhead{\textit{K}}\\
\colhead{} & \colhead{(mag)} &\colhead{(mag)}& \colhead{(mag)}& \colhead{(mag)}&\colhead{(mag)}& \colhead{(mag)}& \colhead{(mag)}& \colhead{(mag)}& \colhead{(mag)}& \colhead{(mag)}
}
\startdata
1	&15.16(01)	&14.60(03)	&15.96(01)	&14.83(01)	&14.52(01)	&14.43(01)	&14.42(01)	&13.58(02)	&13.30(02)	&13.29(03)	\\
2	&13.82(03)	&13.26(03)	&14.86(04)	&13.49(01)	&13.29(01)	&13.13(01)	&13.03(01)	&12.12(02)	&11.87(02)	&11.81(02)	\\
3	&15.92(02)	&15.26(03)	&16.98(04)	&15.53(01)	&15.13(01)	&15.01(01)	&14.99(01)	&14.13(03)	&13.81(04)	&13.78(04)	\\
4	&15.84(03)	&15.10(03)	&16.43(20)	&16.04(01)	&15.03(01)	&14.86(01)	&14.82(01)	&13.78(02)	&13.20(03)	&13.14(03)	\\
5	&15.95(03)	&15.18(03)	&17.08(01)	&15.50(01)	&14.31(01)	&14.19(01)	&14.17(01)	&13.89(03)	&13.48(03)	&13.38(04)	\\
6	&15.02(01)	&14.42(03)	&15.79(04)	&14.66(01)	&14.64(01)	&14.48(01)	&14.42(01)	&13.30(02)	&12.98(02)	&12.94(03)	\\
7	&15.51(03)	&14.78(03)	&16.55(02)	&15.09(01)	&13.21(01)	&13.10(01)	&13.11(01)	&13.47(02)	&13.10(03)	&13.11(03)	\\
8	&14.17(03)	&13.40(01)	&15.23(03)	&13.71(01)	&12.83(01)	&12.53(01)	&12.36(01)	&12.18(02)	&11.88(02)	&11.87(02)	\\
\enddata
\tablecomments{Uncertainties are enclosed in parentheses and are $1\sigma$, in units of 0.01 mag.}
\end{deluxetable*}

\startlongtable
\begin{deluxetable*}{lcccccccccc}
\tablecaption{$BVugriz-$ band Photometry of SN 2022erq
\label{tab:photo}}
\tabletypesize{\scriptsize}
\tablewidth{0pt}
\tablehead{
\colhead{MJD} &  \colhead{$\tau^a$} & \colhead{$t^b$}&\colhead{\textit{B}} & \colhead{\textit{V}}& \colhead{\textit{u}} &\colhead{\textit{g}}& \colhead{\textit{r}}& \colhead{\textit{i}}& \colhead{\textit{z}} & \colhead{Telescope}\\
\colhead{} &\colhead{(days)}& \colhead{(days)} &\colhead{(mag)}& \colhead{(mag)}& \colhead{(mag)}&\colhead{(mag)}& \colhead{(mag)}& \colhead{(mag)}& \colhead{(mag)}& \colhead{}
}
\startdata
59650.93	&1.83&-23.77	&\nodata&\nodata&17.87(09)	&17.75(01)	&17.60(02)	&17.96(02)	&17.39(06)	&LJT	\\
59651.94	&2.84&-22.76	&\nodata&\nodata&\nodata&17.50(02)	&17.40(02)	&17.65(03)	&\nodata&WMS	\\
59652.88	&3.78&-21.82	&\nodata&\nodata&17.43(12)	&17.29(03)	&17.26(03)	&17.48(03)	&17.25(05)	&WMS	\\
59654.17	&5.07&-20.53	&17.34(07)	&16.88(20)	&17.14(08)	&\nodata&\nodata&\nodata&\nodata&ST	\\
59654.93	&5.83&-19.77	&\nodata&\nodata&\nodata&17.04(02)	&16.99(02)	&17.26(03)	&\nodata&WMS	\\
59655.92	&6.82&-18.78	&\nodata&\nodata&17.04(05)	&16.88(03)	&17.02(03)	&17.14(03)	&17.07(06)	&LJT	\\
59655.95	&6.85&-18.75	&\nodata&\nodata&\nodata&16.85(04)	&16.94(04)	&17.06(08)	&\nodata&WMS	\\
59656.93	&7.83&-17.77	&\nodata&\nodata&\nodata&16.80(03)	&16.88(02)	&17.02(02)	&\nodata&LJT	\\
59656.93	&7.83&-17.77	&\nodata&\nodata&\nodata&16.81(02)	&16.82(02)	&16.99(02)	&\nodata&WMS	\\
59657.91	&8.81&-16.79	&\nodata&\nodata&\nodata&16.78(01)	&16.78(02)	&16.93(03)	&\nodata&WMS	\\
59658.84	&9.74&-15.86	&16.96(03)	&16.73(03)	&\nodata&16.74(01)	&16.76(03)	&16.95(04)	&\nodata&TNT	\\
59658.84	&9.74&-15.86	&\nodata&\nodata&\nodata&16.69(03)	&16.70(02)	&16.85(03)	&\nodata&WMS	\\
59659.86	&10.76&-14.84	&16.88(03)	&16.65(03)	&\nodata&16.72(01)	&16.67(03)	&16.87(03)	&\nodata&TNT	\\
59659.87	&10.77&-14.83	&\nodata&\nodata&\nodata&16.61(04)	&16.62(03)	&16.82(02)	&\nodata&LJT	\\
59659.90	&10.80&-14.80	&\nodata&\nodata&\nodata&16.60(02)	&16.64(02)	&16.72(02)	&\nodata&WMS	\\
59660.84	&11.74&-13.86	&\nodata&\nodata&\nodata&16.47(02)	&16.60(02)	&16.71(02)	&\nodata&WMS	\\
59661.10	&12.00&-13.60	&16.73(02)	&16.53(01)	&16.91(07)	&16.57(01)	&16.60(01)	&16.64(01)	&\nodata&ST	\\
59661.89	&12.79&-12.81	&\nodata&\nodata&\nodata&16.39(04)	&16.55(04)	&16.64(06)	&\nodata&WMS	\\
59662.84	&13.74&-11.86	&\nodata&\nodata&\nodata&16.39(01)	&16.48(02)	&16.58(02)	&\nodata&WMS	\\
59663.16	&14.06&-11.54	&16.69(03)	&\nodata&16.85(07)	&16.53(01)	&16.52(01)	&\nodata&16.74(02)	&LT	\\
59663.83	&14.73&-10.87	&\nodata&\nodata&\nodata&16.51(04)	&16.45(04)	&16.55(06)	&\nodata&WMS	\\
59664.16	&15.06&-10.54	&\nodata&16.44(02)	&16.79(05)	&16.54(03)	&16.50(02)	&\nodata&16.68(02)	&LT	\\
59665.88	&16.78&-8.82	&\nodata&\nodata&\nodata&16.34(01)	&16.41(02)	&16.48(02)	&\nodata&WMS	\\
59665.89	&16.79&-8.81	&16.69(06)	&16.43(05)	&\nodata&\nodata&\nodata&\nodata&\nodata&TNT	\\
59667.88	&18.78&-6.82	&\nodata&\nodata&16.79(09)	&16.32(01)	&16.37(01)	&16.40(02)	&\nodata&WMS	\\
59668.17	&19.07&-6.53	&16.61(02)	&16.31(02)	&16.83(06)	&16.43(01)	&16.37(01)	&\nodata&\nodata&LT	\\
59668.86	&19.76&-5.84	&\nodata&\nodata&\nodata&16.32(01)	&16.34(01)	&16.37(02)	&\nodata&WMS	\\
59669.17	&20.07&-5.53	&\nodata&16.33(02)	&16.92(05)	&16.43(01)	&16.36(01)	&16.43(01)	&16.59(02)	&LT	\\
59672.86	&23.76&-1.84	&16.59(03)	&16.30(03)	&\nodata&16.43(01)	&16.28(03)	&16.38(02)	&\nodata&TNT	\\
59675.78	&26.68&1.08	&16.57(03)	&16.31(02)	&\nodata&16.45(02)	&16.27(03)	&16.35(03)	&\nodata&TNT	\\
59677.06	&27.96&2.36	&\nodata&\nodata&17.04(06)	&16.37(01)	&16.20(01)	&16.24(01)	&\nodata&ST	\\
59678.90	&29.80&4.20	&\nodata&\nodata&17.08(05)	&16.30(01)	&16.10(01)	&16.24(01)	&\nodata&LJT	\\
59680.06	&30.96&5.36	&\nodata&\nodata&17.08(07)	&16.34(01)	&16.17(01)	&16.20(01)	&\nodata&ST	\\
59681.20	&32.10&6.50	&\nodata&\nodata&\nodata&\nodata&\nodata&\nodata&16.39(01)	&LT	\\
59683.08	&33.98&8.38	&\nodata&\nodata&\nodata&16.32(01)	&16.17(01)	&16.19(01)	&\nodata&ST	\\
59683.10	&34.00&8.40	&\nodata&\nodata&17.18(05)	&\nodata&\nodata&\nodata&16.37(01)	&LT	\\
59684.80	&35.70&10.10	&\nodata&\nodata&17.26(05)	&16.36(01)	&\nodata&16.23(02)	&\nodata&LJT	\\
59685.79	&36.69&11.09	&16.69(03)	&16.27(03)	&\nodata&16.47(01)	&\nodata&\nodata&\nodata&TNT	\\
59687.06	&37.96&12.36	&\nodata&\nodata&\nodata&16.43(01)	&16.22(01)	&16.15(01)	&\nodata&ST	\\
59687.79	&38.69&13.09	&16.77(04)	&16.31(03)	&\nodata&16.52(04)	&16.22(05)	&16.31(03)	&\nodata&TNT	\\
59688.85	&39.75&14.15	&16.82(02)	&16.35(03)	&\nodata&16.51(01)	&16.22(03)	&16.30(03)	&\nodata&TNT	\\
59691.13	&42.03&16.43	&16.77(04)	&16.32(02)	&17.42(08)	&16.52(02)	&16.24(01)	&16.22(01)	&16.42(01)	&LT	\\
59691.77	&42.67&17.07	&\nodata&\nodata&\nodata&16.53(02)	&16.29(02)	&16.25(01)	&\nodata&LJT	\\
59691.80	&42.70&17.10	&16.90(03)	&16.37(02)	&\nodata&16.58(02)	&16.24(03)	&16.30(03)	&\nodata&TNT	\\
59692.80	&43.70&18.10	&16.94(03)	&16.40(03)	&\nodata&16.63(01)	&16.28(03)	&16.30(03)	&\nodata&TNT	\\
59692.88	&43.78&18.18	&\nodata&\nodata&\nodata&16.54(02)	&16.37(01)	&16.14(03)	&\nodata&LJT	\\
59693.10	&44.00&18.40	&16.86(03)	&\nodata&17.46(08)	&16.62(02)	&16.25(01)	&16.22(01)	&16.37(01)	&LT	\\
59696.73	&47.63&22.03	&16.97(04)	&16.43(03)	&\nodata&16.64(02)	&16.31(01)	&16.25(02)	&\nodata&LJT	\\
59697.08	&47.98&22.38	&\nodata&\nodata&\nodata&\nodata&\nodata&\nodata&16.44(02)	&LT	\\
59697.09	&47.99&22.39	&\nodata&\nodata&\nodata&16.68(01)	&16.29(01)	&16.25(01)	&\nodata&ST	\\
59699.79	&50.69&25.09	&17.20(03)	&16.54(03)	&\nodata&16.80(01)	&16.37(03)	&16.39(03)	&\nodata&TNT	\\
59700.80	&51.70&26.10	&17.27(06)	&16.56(03)	&\nodata&16.83(02)	&16.37(03)	&16.39(03)	&\nodata&TNT	\\
59701.16	&52.06&26.46	&17.13(03)	&16.53(02)	&17.77(06)	&16.80(01)	&16.38(01)	&16.31(01)	&16.49(02)	&LT	\\
59701.78	&52.68&27.08	&17.31(05)	&16.55(03)	&\nodata&16.85(02)	&16.37(03)	&16.34(03)	&\nodata&TNT	\\
59702.78	&53.68&28.08	&17.24(04)	&16.55(03)	&\nodata&16.86(02)	&16.37(03)	&\nodata&\nodata&TNT	\\
59705.12	&56.02&30.42	&17.28(02)	&16.63(02)	&17.78(04)	&16.93(01)	&16.44(01)	&16.37(01)	&16.50(01)	&LT	\\
59706.77	&57.67&32.07	&17.36(07)	&16.66(05)	&17.97(09)	&16.97(02)	&16.48(01)	&16.36(07)	&\nodata&LJT	\\
59709.05	&59.95&34.35	&\nodata&\nodata&\nodata&\nodata&\nodata&\nodata&16.50(03)	&LT	\\
59710.06	&60.96&35.36	&\nodata&\nodata&\nodata&17.10(01)	&16.49(01)	&16.39(01)	&\nodata&ST	\\
59710.99	&61.89&36.29	&\nodata&16.77(02)	&\nodata&\nodata&\nodata&\nodata&\nodata&ST	\\
59714.76	&65.66&40.06	&17.70(03)	&16.88(03)	&\nodata&17.24(02)	&16.54(04)	&16.55(03)	&\nodata&TNT	\\
59716.06	&66.96&41.36	&\nodata&\nodata&\nodata&\nodata&\nodata&\nodata&16.62(01)	&LT	\\
59716.76	&67.66&42.06	&17.72(03)	&16.89(03)	&\nodata&17.36(03)	&16.60(03)	&16.61(03)	&\nodata&TNT	\\
59719.06	&69.96&44.36	&17.78(02)	&16.94(02)	&\nodata&17.27(01)	&16.67(01)	&16.49(01)	&\nodata&ST	\\
59719.79	&70.69&45.09	&17.82(02)	&17.01(02)	&\nodata&17.41(01)	&16.71(03)	&16.67(02)	&\nodata&TNT	\\
59720.70	&71.60&46.00	&17.87(04)	&16.99(03)	&\nodata&17.36(02)	&16.68(03)	&16.69(03)	&\nodata&TNT	\\
59721.68	&72.58&46.98	&17.97(03)	&17.03(02)	&\nodata&17.38(03)	&16.75(04)	&16.63(03)	&\nodata&TNT	\\
59722.68	&73.58&47.98	&17.88(03)	&17.05(02)	&\nodata&17.40(02)	&16.73(03)	&16.72(03)	&\nodata&TNT	\\
59725.62	&76.52&50.92	&18.04(03)	&17.10(03)	&\nodata&17.50(02)	&16.78(03)	&16.70(03)	&\nodata&TNT	\\
59727.05	&77.95&52.35	&18.04(02)	&17.09(01)	&\nodata&17.49(01)	&16.82(01)	&16.63(01)	&\nodata&ST	\\
59727.71	&78.61&53.01	&18.04(03)	&17.13(02)	&\nodata&17.56(02)	&16.87(03)	&16.69(03)	&\nodata&TNT	\\
59728.73	&79.63&54.03	&18.03(03)	&17.16(02)	&\nodata&17.59(02)	&16.84(03)	&16.83(03)	&\nodata&TNT	\\
59729.16	&80.06&54.46	&\nodata&\nodata&\nodata&\nodata&\nodata&\nodata&16.81(01)	&LT	\\
59730.75	&81.65&56.05	&18.12(04)	&17.22(02)	&\nodata&17.59(02)	&16.88(03)	&16.85(03)	&\nodata&TNT	\\
59731.97	&82.87&57.27	&\nodata&\nodata&\nodata&\nodata&\nodata&\nodata&16.80(02)	&LT	\\
59732.76	&83.66&58.06	&18.02(03)	&17.22(02)	&\nodata&17.62(02)	&16.95(02)	&16.83(02)	&\nodata&LJT	\\
59735.05	&85.95&60.35	&\nodata&\nodata&\nodata&17.67(01)	&16.95(01)	&16.77(02)	&\nodata&ST	\\
59736.64	&87.54&61.94	&18.29(05)	&17.33(04)	&\nodata&17.72(03)	&17.04(04)	&16.95(03)	&\nodata&TNT	\\
59742.07	&92.97&67.37	&\nodata&\nodata&\nodata&17.72(01)	&17.12(01)	&\nodata&\nodata&ST	\\
59743.03	&93.93&68.33	&18.32(03)	&17.33(02)	&\nodata&\nodata&\nodata&16.89(01)	&\nodata&ST	\\
59744.94	&95.84&70.24	&\nodata&\nodata&\nodata&\nodata&\nodata&\nodata&16.95(03)	&LT	\\
59748.73	&99.63&74.03	&18.28(03)	&17.37(02)	&\nodata&17.77(02)	&17.20(02)	&17.07(01)	&\nodata&LJT	\\
59754.99	&105.89&80.29	&18.36(02)	&17.48(01)	&\nodata&17.89(01)	&17.28(01)	&17.11(01)	&\nodata&ST	\\
59768.01	&118.91&93.31	&\nodata&\nodata&\nodata&\nodata&\nodata&\nodata&17.24(02)	&LT	\\
59768.96	&119.86&94.26	&18.59(02)	&17.71(02)	&\nodata&17.98(01)	&17.51(01)	&17.33(01)	&\nodata&ST	\\
59772.98	&123.88&98.28	&18.88(23)	&17.87(17)	&\nodata&18.01(16)	&17.50(14)	&17.42(14)	&17.38(29)	&KOT	\\
59775.99	&126.89&101.29	&18.82(21)	&17.86(15)	&\nodata&18.07(16)	&17.49(14)	&17.36(13)	&17.32(17)	&KOT	\\
59778.95	&129.85&104.25	&18.63(16)	&17.84(12)	&\nodata&18.18(12)	&17.60(10)	&17.44(10)	&17.42(12)	&KOT	\\
59781.94	&132.84&107.24	&18.99(17)	&17.92(09)	&\nodata&18.23(11)	&17.66(07)	&17.57(06)	&17.64(10)	&KOT	\\
59784.93	&135.83&110.23	&18.97(19)	&17.95(10)	&\nodata&18.26(13)	&17.63(09)	&17.61(07)	&17.41(11)	&KOT	\\
59786.96	&137.86&112.26	&\nodata&\nodata&\nodata&\nodata&\nodata&\nodata&17.57(03)	&LT	\\
59788.92	&139.82&114.22	&18.91(18)	&18.01(10)	&\nodata&18.32(12)	&17.72(08)	&17.62(06)	&17.47(11)	&KOT	\\
59792.91	&143.81&118.21	&18.94(17)	&18.10(10)	&\nodata&18.43(11)	&17.79(08)	&17.64(06)	&17.66(12)	&KOT	\\
59794.02	&144.92&119.32	&\nodata&17.97(02)	&\nodata&18.32(01)	&17.77(01)	&17.61(01)	&\nodata&ST	\\
59795.90	&146.80&121.20	&19.05(17)	&18.06(08)	&\nodata&18.40(10)	&17.82(07)	&17.64(06)	&17.61(12)	&KOT	\\
59797.88	&148.78&123.18	&\nodata&\nodata&\nodata&\nodata&\nodata&\nodata&17.65(02)	&LT	\\
59798.89	&149.79&124.19	&\nodata&\nodata&\nodata&\nodata&\nodata&\nodata&17.71(03)	&LT	\\
59798.92	&149.82&124.22	&19.05(27)	&18.08(13)	&\nodata&18.58(15)	&17.81(09)	&17.74(08)	&17.75(16)	&KOT	\\
59801.88	&152.78&127.18	&18.94(21)	&18.04(11)	&\nodata&18.47(14)	&18.00(08)	&17.75(06)	&17.70(15)	&KOT	\\
59805.87	&156.77&131.17	&19.19(45)	&18.26(15)	&\nodata&18.62(15)	&18.04(12)	&17.85(11)	&17.80(16)	&KOT	\\
59808.89	&159.79&134.19	&19.24(21)	&18.25(13)	&\nodata&18.49(13)	&17.99(11)	&17.81(10)	&17.93(18)	&KOT	\\
59816.86	&167.76&142.16	&\nodata&\nodata&\nodata&\nodata&\nodata&\nodata&17.90(02)	&LT	\\
59816.88	&167.78&142.18	&19.24(20)	&18.46(12)	&\nodata&18.67(11)	&18.17(10)	&18.02(10)	&17.97(17)	&KOT	\\
59817.53	&168.43&142.83	&19.09(05)	&18.25(04)	&\nodata&18.69(03)	&18.03(03)	&18.09(04)	&\nodata&TNT	\\
59822.57	&173.47&147.87	&19.19(05)	&18.30(04)	&\nodata&18.66(03)	&18.04(03)	&18.11(04)	&\nodata&TNT	\\
59823.54	&174.44&148.84	&19.10(06)	&18.33(04)	&\nodata&18.69(02)	&18.08(04)	&18.12(03)	&\nodata&TNT	\\
59823.87	&174.77&149.17	&19.25(22)	&18.46(16)	&\nodata&18.71(17)	&18.34(18)	&17.88(15)	&17.82(24)	&KOT	\\
59824.69	&175.59&149.99	&19.21(10)	&18.44(06)	&\nodata&\nodata&\nodata&\nodata&\nodata&TNT	\\
59825.98	&176.88&151.28	&\nodata&\nodata&\nodata&\nodata&\nodata&\nodata&17.92(03)	&LT	\\
59827.00	&177.90&152.30	&\nodata&18.44(03)	&\nodata&18.68(02)	&18.29(02)	&18.11(02)	&\nodata&ST	\\
59827.60	&178.50&152.90	&19.06(04)	&18.37(04)	&\nodata&18.79(05)	&18.17(03)	&18.23(03)	&\nodata&TNT	\\
59828.59	&179.49&153.89	&19.11(06)	&18.36(04)	&\nodata&18.77(04)	&18.11(03)	&18.19(04)	&\nodata&TNT	\\
59829.60	&180.50&154.90	&19.21(07)	&18.41(04)	&\nodata&18.79(06)	&18.29(04)	&18.24(04)	&\nodata&TNT	\\
59834.90	&185.80&160.20	&19.33(59)	&18.51(19)	&\nodata&18.76(17)	&18.46(13)	&18.06(13)	&18.01(34)	&KOT	\\
59838.85	&189.75&164.15	&\nodata&\nodata&\nodata&\nodata&\nodata&\nodata&18.10(05)	&LT	\\
59839.93	&190.83&165.23	&19.52(04)	&18.53(02)	&\nodata&18.82(02)	&18.49(02)	&18.12(02)	&\nodata&ST	\\
59840.60	&191.50&165.90	&19.28(05)	&18.43(04)	&\nodata&18.80(03)	&18.25(03)	&18.25(04)	&\nodata&TNT	\\
59845.79	&196.69&171.09	&19.54(20)	&18.77(15)	&\nodata&19.02(15)	&18.48(13)	&18.24(13)	&18.25(22)	&KOT	\\
59853.51	&204.41&178.81	&19.49(06)	&18.69(03)	&\nodata&19.02(03)	&18.60(03)	&18.34(03)	&\nodata&LJT	\\
59856.59	&207.49&181.89	&19.33(09)	&18.62(07)	&\nodata&18.97(05)	&18.33(04)	&18.20(07)	&\nodata&TNT	\\
59856.93	&207.83&182.23	&19.74(33)	&18.91(18)	&\nodata&19.16(17)	&18.49(12)	&18.53(14)	&\nodata&KOT	\\
59858.57	&209.47&183.87	&19.41(10)	&18.64(05)	&\nodata&18.94(07)	&18.45(04)	&18.40(05)	&\nodata&TNT	\\
59859.85	&210.75&185.15	&19.48(11)	&18.78(05)	&\nodata&\nodata&18.64(04)	&18.49(03)	&18.24(04)	&LT	\\
59861.53	&212.43&186.83	&19.51(15)	&18.88(12)	&\nodata&19.03(12)	&18.43(06)	&18.41(05)	&\nodata&TNT	\\
59862.58	&213.48&187.88	&19.59(19)	&18.73(10)	&\nodata&19.07(12)	&18.42(06)	&\nodata&\nodata&TNT	\\
59868.89	&219.79&194.19	&19.70(34)	&18.96(19)	&\nodata&19.11(21)	&18.59(13)	&18.63(16)	&18.42(14)	&KOT	\\
59869.91	&220.81&195.21	&19.70(06)	&18.82(04)	&\nodata&19.09(03)	&18.69(03)	&\nodata&18.26(05)	&LT	\\
59872.56	&223.46&197.86	&19.66(09)	&18.75(04)	&\nodata&19.12(04)	&18.51(03)	&18.49(05)	&\nodata&LJT	\\
59873.50	&224.40&198.80	&\nodata&18.81(02)	&\nodata&\nodata&18.75(02)	&18.61(03)	&18.26(04)	&LJT	\\
59879.49	&230.39&204.79	&19.79(06)	&18.86(03)	&\nodata&19.15(03)	&18.89(02)	&18.63(03)	&\nodata&LJT	\\
59879.85	&230.75&205.15	&19.91(23)	&18.99(15)	&\nodata&19.15(13)	&18.69(10)	&18.48(10)	&18.45(24)	&KOT	\\
59885.50	&236.40&210.80	&\nodata&18.89(03)	&\nodata&\nodata&18.83(03)	&18.72(03)	&18.34(04)	&LJT	\\
59893.49	&244.39&218.79	&19.89(09)	&19.03(03)	&\nodata&19.33(04)	&18.93(03)	&18.77(03)	&18.39(05)	&LJT	\\
59896.48	&247.38&221.78	&20.20(10)	&19.01(05)	&\nodata&19.24(03)	&19.01(03)	&18.83(04)	&18.45(04)	&LJT	\\
59896.69	&247.59&221.99	&19.87(23)	&19.02(18)	&\nodata&19.30(17)	&18.80(13)	&18.46(13)	&18.43(21)	&KOT	\\
59898.49	&249.39&223.79	&20.01(05)	&18.99(04)	&\nodata&19.27(02)	&18.93(03)	&18.76(03)	&\nodata&LJT	\\
59898.86	&249.76&224.16	&19.69(09)	&19.11(06)	&\nodata&19.52(06)	&19.15(06)	&18.82(03)	&\nodata&LT	\\
59903.50	&254.40&228.80	&20.05(06)	&19.14(04)	&\nodata&19.45(04)	&19.18(04)	&18.81(03)	&18.53(05)	&LJT	\\
59913.48	&264.38&238.78	&\nodata&19.17(06)	&\nodata&\nodata&19.06(05)	&18.73(04)	&\nodata&LJT	\\
59920.48	&271.38&245.78	&\nodata&\nodata&\nodata&\nodata&19.35(12)	&19.00(10)	&\nodata&LJT	\\
59941.19	&292.09&266.49	&20.18(48)	&19.36(39)	&\nodata&19.61(33)	&19.15(20)	&19.07(26)	&\nodata&KOT	\\
59969.16	&320.06&294.46	&20.40(28)	&19.38(20)	&\nodata&19.83(19)	&19.31(14)	&19.23(12)	&19.24(23)	&KOT	\\
59981.09	&331.99&306.39	&\nodata&\nodata&\nodata&\nodata&\nodata&19.46(26)	&19.43(46)	&KOT	\\
59986.08	&336.98&311.38	&\nodata&19.68(38)	&\nodata&19.93(30)	&19.44(20)	&19.43(24)	&19.40(23)	&KOT	\\
59991.05	&341.95&316.35	&\nodata&\nodata&\nodata&19.92(33)	&19.52(21)	&19.48(20)	&19.46(26)	&KOT	\\
59992.94	&343.84&318.24	&\nodata&\nodata&\nodata&\nodata&19.68(05)	&19.52(03)	&\nodata&LJT	\\
59993.91	&344.81&319.21	&\nodata&\nodata&\nodata&19.99(14)	&\nodata&\nodata&\nodata&LJT	\\
60005.05	&355.95&330.35	&20.82(48)	&\nodata&\nodata&20.29(50)	&19.89(32)	&19.59(30)	&19.57(13)	&KOT	\\
60013.07	&363.97&338.37	&\nodata&20.34(13)	&\nodata&\nodata&20.09(46)	&19.63(41)	&\nodata&KOT	\\
60019.97	&370.87&345.27	&\nodata&\nodata&\nodata&\nodata&20.59(29)	&20.19(40)	&\nodata&KOT	\\
60024.98	&375.88&350.28	&21.56(35)	&20.78(41)	&\nodata&21.03(38)	&20.63(21)	&20.23(17)	&\nodata&KOT	\\
60032.01	&382.91&357.31	&21.75(53)	&\nodata&\nodata&\nodata&20.82(73)	&20.41(45)	&\nodata&KOT	\\
60049.86 & 400.76 & 375.16 & \nodata & \nodata & \nodata & \nodata & 20.69(10) & 20.47(39) & 20.12(19) & LJT \\
60052.90 & 403.80 & 378.20 & \nodata & \nodata & \nodata & \nodata & 20.93(16) & 20.61(11) & \nodata    & LJT \\
60087.84 & 438.74 & 413.14 & 22.53(30) & 21.65(26) & \nodata & 22.03(20) & 21.62(25) & 21.37(23) & \nodata & LJT \\
60091.88 & 442.78 & 417.18 & \nodata & \nodata & \nodata & 22.16(15) & 21.66(25) & 21.47(36) & \nodata & LJT \\
60092.88 & 443.78 & 418.18 & \nodata & \nodata & \nodata & \nodata & 22.43(30) & 22.05(49) & \nodata & LJT \\
60125.60 & 476.50 & 450.90 & \nodata & \nodata & \nodata & \nodata & 23.15(43) & \nodata    & \nodata & LJT \\
\enddata
\tablecomments{Uncertainties are enclosed in parentheses and are $1\sigma$, in units of 0.01 mag. }
$^a${Phase relative to the day of explosion, MJD = 59649.1.}\\
$^b${Phase relative to the day of $B$-band maximum, MJD = 59674.7.}

\end{deluxetable*}

\begin{deluxetable*}{lccccc}[!th]
\label{app_a:JHK}
\tablecaption{$JHK-$ band Photometry of SN 2022erq obtained by NOTCam of NOT}
\tabletypesize{\scriptsize}
\tablewidth{0pt}
\tablehead{
\colhead{MJD}  & \colhead{$\tau^a$}&\colhead{t$^b$}& \colhead{\textit{J}} & \colhead{\textit{H}}& \colhead{\textit{K}}\\
\colhead{} & \colhead{(days)} & \colhead{(days)} &\colhead{(mag)}& \colhead{(mag)}& \colhead{(mag)}
}
\startdata
59683.70	& 	34.60	& 	9.00	& 	15.73(01)	& 	15.52(01)	& 	15.51(02)	\\
59809.70	& 	160.60	 &	135.00	 &	17.34(01)	 &	16.97(04)	 &	16.88(02)	 \\
59837.64	& 	188.54	 &	162.94	 &	17.64(03)	 &	17.35(03)	 &	17.21(03)	 \\
59882.47	& 	233.37	 &	207.77	 &	17.92(04)	 &	17.59(03)	 &	17.49(04)	 \\
59997.86	& 	348.76	 &	323.16	 &	18.21(09)	 &	18.05(05)	 &	17.49(08)	 \\
60018.77	& 	369.67	 &	344.07	 &	18.68(08)	 &	18.18(07)	 &	17.58(05)	 \\
60060.71	& 	411.61	 &	386.01	 &	18.94(05)	 &	18.32(08)	 &	17.67(07)	 \\
\enddata
\tablecomments{Uncertainties are enclosed in parentheses and are $1\sigma$, in units of 0.01 mag. }
$^a${Phase relative to the day of explosion, MJD = 59649.1.}\\
$^b${Phase relative to the day of $B$-band maximum, MJD = 59674.7.}
\end{deluxetable*}

\begin{table*}
\caption{Journal of Spectroscopic Observations of SN 2022erq}
\scriptsize
\centering
\begin{tabular}{cccccc}
\hline\hline
Date (YMD) & MJD & Epoch$^a$ (d) & Range (\AA) & $R^b$ & Telescope+Inst.\\
\hline
220312 & 59650.902 & 1.8 & 3488-8742 & 360 & LJT+YFOSC \\
220315 & 59653.904 & 4.8 & 3498-8767 & 320 & LJT+YFOSC \\
220320 & 59658.850 & 9.7 & 3875-8854 & 300 & XLT+BFOSC \\
220321 & 59659.843 & 11 & 3486-8742 & 420 & LJT+YFOSC \\
220324 & 59662.517 & 13 & 3620-10670 & 470 & Shane+KDS \\
220330 & 59668.219 & 19 & 3425-9701 & 430 & NOT+ALFOSC \\
220330 & 59668.239 & 19 & 6134-7772 & 1420 & TNG+VHR \\
220409 & 59678.873 & 30 & 3486-8742 & 400 & LJT+YFOSC \\
220411 & 59680.182 & 31 & 3410-9665 & 350 & NOT+ALFOSC \\
220415 & 59684.777 & 36 & 3503-8768 & 300 & LJT+YFOSC \\
220421 & 59691.147 & 42 & 3500-10318 & 570 & TNG+LRB/LRR \\
220424 & 59693.850 & 45 & 3497-8484 & 300 & LJT+YFOSC \\
220512 & 59711.686 & 63 & 3862-8831 & 370 & XLT+BFOSC \\
220512 & 59711.709 & 63 & 3474-8766 & 320 & LJT+YFOSC \\
220513 & 59712.723 & 64 & 3862-8831 & 380 & XLT+BFOSC \\
220520 & 59719.746 & 71 & 3874-8839 & 410 & XLT+BFOSC \\
220526 & 59726.148 & 77 & 6139-7774 & 1420 & TNG+VHR \\
220618 & 59748.085 & 99 & 3422-10071 & 410 & NOT+ALFOSC \\
220712 & 59772.122 & 123 & 3424-9700 & 450 & NOT+ALFOSC \\
220805 & 59796.399 & 147 & 3623-8107 & 1620 & Shane+KDS \\
220808 & 59799.064 & 150 & 8960-13296 & \nodata & GTC+EMIR \\
220826 & 59817.614 & 169 & 3897-8876 & 370 & XLT+BFOSC \\
220917 & 59839.468 & 190 & 3891-8869 & 380 & XLT+BFOSC \\
220925 & 59847.189 & 198 & 3622-9998 & 550 & Shane+KDS \\
221018 & 59870.594 & 221 & 3896-8868 & 360 & XLT+BFOSC \\
221027 & 59879.524 & 230 & 3520-8784 & 350 & LJT+YFOSC \\
221109 & 59892.489 & 243 & 3521-8783 & 400 & LJT+YFOSC \\
230301 & 60004.890 & 356 & 3631-8947 & 300 & LJT+YFOSC \\
230316 & 60019.539 & 370 & 3622-10000 & 500 & Shane+KDS \\
230322 & 60025.619 & 377 & 4500-9650 & 1550 & KeckII+DEIMOS \\
230709 & 60137.280 & 488 & 3650-10000 & 1800 & HET+LRS2 \\
251108 & 60987.490 & 1338 & 3850-7420 & 630 & LJT+YFOSC \\
251116 & 60995.480 & 1346 & 3643-8863 & 350 & LJT+YFOSC \\
\hline\hline
\end{tabular}

\raggedright
$^a${The epoch is relative to the derived explosion date, MJD = 59649.1.}\\
$^b${Spectral resolving power $R = \lambda / \Delta\lambda$, where $\lambda = 6300$ \AA; $\Delta\lambda$ is derived by the FWHM of the night-sky line [O~I] $\lambda$6300.3, except for the NIR spectrum from GTC.}\\

\label{Tab:Spec_log}
\end{table*}

\clearpage

\bibliographystyle{aasjournalv7}

\end{document}